\documentclass[%
 reprint,
 amsmath,amssymb,
 aps,
]{revtex4-1}
\usepackage{braket}
\usepackage{graphicx}
\usepackage{dcolumn}
\usepackage{bm}
\usepackage {multirow}
\usepackage{float}
\usepackage{lipsum}
\usepackage{amsmath}
\usepackage{color}

\begin{document}
\makeatletter
\newcommand*{\rom}[1]{\expandafter\@slowromancap\romannumeral #1@}
\makeatother

\preprint{APS/123-QED}

\title{Nuclear spin relaxation rate near the disorder-driven quantum critical point
\\
in Weyl fermion systems 
}

\author{Tomoki Hirosawa, Hideaki Maebashi, and Masao Ogata}

\affiliation{%
 Department of Physics, University of Tokyo, Bunkyo, Tokyo 113-0033, Japan 
}%

\date{\today}

\begin{abstract}
Disorder such as impurities and dislocations in Weyl semimetals (SMs)
drives a quantum critical point (QCP) 
where the density of states at the Weyl point gains a non-zero value. 
Near the QCP, the asymptotic low energy singularities of physical quantities 
are controlled by the critical exponents $\nu$ and $z$.  
The nuclear spin-lattice relaxation rate, which originates from the hyperfine coupling between 
a nuclear spin and long-range orbital currents in Weyl fermion systems, shows intriguing critical behavior.
Based on the self-consistent Born approximation for impurities, 
we study the nuclear spin-lattice relaxation rate $1/T_1$ 
due to the orbital currents in disordered Weyl SMs.
We find that $(T_1T)^{-1}\sim E^{2/z}$ at the QCP where $E$ is the maximum of temperature $T$ and chemical potential $\mu(T)$ relative to the Weyl point.
This scaling behavior of $(T_1T)^{-1}$ is also confirmed by the self-consistent $T$-matrix approximation, where a remarkable temperature dependence of 
$\mu(T)$ could play an important role. 
We hope these results of $(T_1T)^{-1}$ will serve as an impetus for exploration of the 
disorder-driven quantum criticality in Weyl materials.
\end{abstract}

\maketitle


\section{Introduction}
\label{Introduction}

In condensed matter physics, the Weyl Hamiltonian describes an effective model of gapless systems where the inversion or time-reversal symmetry is broken, known as a Weyl semimetal (SM)~\cite{armitage}. 
Among many candidates, a family of TaAs-type materials is a typical example of Weyl SMs~\cite{huang,weng,xu,lv,shekhar}. 
Since the discovery of these materials, the unusual galvanomagnetic transport has been attracted much attention.
When the electric and magnetic field is applied in parallel, the negative magnetoresistance was predicted due to the chiral anomaly~\cite{son,burkov}. 
In TaAs-type Weyl SMs, the negative magnetoresistance was experimentally observed~\cite{arnold,xhuang,zwang,clzhang}.
Similar studies were carried out on thermoelectric transport. 
A large positive contribution proportional to the square of the magnetic field was predicted in the longitudinal thermal conductivity when the temperature gradient and magnetic field is applied in parallel~\cite{lundgren}. 

Besides these transport properties, nuclear magnetic resonance (NMR) in Weyl SMs 
shows unusual dependence of the nuclear spin-lattice relaxation time $T_1$ on temperature $T$. 
In general, the inverse of $T_1T$ detects local fluctuations of a magnetic field 
produced at a nuclear spin site by the surrounding electrons~\cite{abragam}. 
It is usually scaled as the square of the density 
of states, called the Korringa relation~\cite{moriya,korringa}. 
Since the density of states in Weyl SMs is proportional to the square of the energy 
around the Weyl point, a naive power counting based on the Korringa relation expects that 
$(T_1T)^{-1} \sim {\rm max}[\mu(T)^4, T^4]$, 
where $\mu(T)$ is the chemical potential measured from the Weyl point.
However, recent nuclear quadrupole resonance (NQR) experiment on TaP 
revealed that $(T_1T)^{-1} \sim {\rm max}[\mu(T)^2, T^2]$ in Weyl SMs with remarkable temperature dependence of $\mu(T)$~\cite{yasuoka,okvatovity}. 
This unusual scaling had been predicted as an orbital effect 
in $(T_1T)^{-1}$ which originates from the hyperfine coupling between a nuclear spin 
and long-range orbital currents of Weyl fermions~\cite{dora,hirosawa}. 
The importance of temperature dependence of 
$\mu(T)$ in gapless systems was also pointed out in relation to 
the Hall coefficient observed in $\alpha$-(BEDT-TTF)$_2$I$_3$~\cite{akobayashi,tajima,kajita}.

In the past, the orbital magnetism has been studied extensively in Dirac materials, which are narrow-gap electron systems described by the Dirac Hamiltonian~\cite{wolff,fuseya}. 
The large diamagnetism of bismuth-antimony alloys Bi$_{1-x}$Sb$_x$ demonstrated a significant contribution from the interband matrix element of current operator~\cite{wehrli,fukuyama}. Recently, it was shown to be counterpart of the inverse of the charge renormalization factor in quantum electrodynamics~\cite{maebashi1}. 
Furthermore, $(T_1T)^{-1}$ in Dirac electron systems was found to be proportional to $T^2$ due to the orbital effect when temperature is higher than the band gap~\cite{hirosawa}. 
This finding partly explains the temperature dependence of $1/T_1$ observed in the $\beta$-detected NMR experiment on Bi$_{0.9}$Sb$_{0.1}$~\cite{macfarlane}.

The nuclear spin-lattice relaxation rate due to long-range orbital currents was   
formulated by Lee and Nagaosa, considering local fluctuations of 
the Biot-Savart magnetic field produced by the orbital current~\cite{nagaosa}. 
In three dimensions, 
it can be written in terms of the transverse conductivity $\sigma_{\rm T} (q,\omega)$ 
with a wavevector $q$ and frequency $\omega$ as
\begin{align}
\frac{1}{T_1T}&=\frac{4k_\textrm{B}}{3} \gamma_\textrm{n}^2 \mu_0^2 
\int \frac{d^3q}{(2 \pi)^3} 
\frac{1}{q^2}{\rm Re}\,\sigma_\textrm{T}(q,\omega_0), 
\label{Eq. 1}
\end{align}
where $ \gamma_\textrm{n}$, $\mu_0$, and $\omega_0$ are the  gyromagnetic ratio of a nuleus, 
the vacuum permeability, and  the nuclear Larmor frequency, respectively. 
For Dirac electron systems with an electronic charge $-e$, 
effective mass $m^*$, and half band gap $\Delta$, 
Eq.~(\ref{Eq. 1}) leads to~\cite{maebashi2}
\begin{align}
\frac{1}{T_1T}
=& \frac{2 \pi k_\textrm{B} }{3} \gamma_{\rm n}^2 \mu_0^2 e^2 {c^*}^4\hbar^3
\nonumber
\\
&\times
\int_{-\infty}^{\infty} d E 
\left[ - \frac{\partial n_\textrm{F} (E)}{\partial E} \right]
\frac{D^2(E)}{E^2} 
\log \frac{2(E^2 - \Delta^2)}{|E| \omega_0} ,
\label{Eq. 2}
\end{align}
where $n_\textrm{F}(E) = [e^{(E-\mu)/k_\textrm{B}T} + 1]^{-1}$ is the 
Fermi distribution function and $D(E)$ is the density of states. 
In the low electron density limit of $\mu \to \Delta$ with $\mu > \Delta$, 
Eq.~(\ref{Eq. 2}) corresponds to the result for free electron gas~\cite{knigavko}, 
which shows the usual scaling $(T_1T)^{-1}\sim D^2(\mu) \sim \mu - \Delta$ 
(except the logarithmic dependence).
On the other hand, Eq.~(\ref{Eq. 2}) provides the result for Weyl SMs in the gapless limit of $\Delta \to 0$ with fixing $c^* \equiv \sqrt{\Delta/m^*}$, 
which shows the unusual scaling $(T_1T)^{-1} \sim E^2$ 
with $E$ the maximum of $T$ and $\mu$~\cite{dora}. 
We thus see that the origin of the unusual scaling is attributed to the 
gapless structure of the density of states. 
It is, however, known that the density of states in gapless systems 
is sensitive to the presence of disorder such as impurities and dislocations. 
Then a natural question arises: 
{\it How does the scaling behavior of $(T_1T)^{-1}$ change by introducing the disorder?}

The disorder-induced quantum critical point (QCP) exists in $d$-dimensional gapless systems 
with a dispersion $E_k \propto k^{\alpha}$ under the short-ranged disorder ~\cite{fradkin1,fradkin2,sergev_prl,sergev_rev}. 
When $d < 2 \alpha$, 
the gapless structure of the density of states disappears by infinitesimal disorder. 
When $d > 2 \alpha$, 
on the other hand, it is robust against weak disorder below the critical strength, 
leading to the disorder-driven QCP. 
Since $d=3$ and $\alpha=1$ for Weyl SMs, disordered Weyl SMs have the QCP~\cite{sergev_prb,ominato,nandkishore,shindou,shapourian,sbierski,pixley,luo,louvet,hosur,goswami,chen,bera,roy,kobayashi1,kobayashi2,rare_pixley}. 
Here, we should note that the critical behavior is correct only up to rare-event effects~\cite{nandkishore,rare_pixley}.
\begin{figure}[t]
\includegraphics[width=75mm]{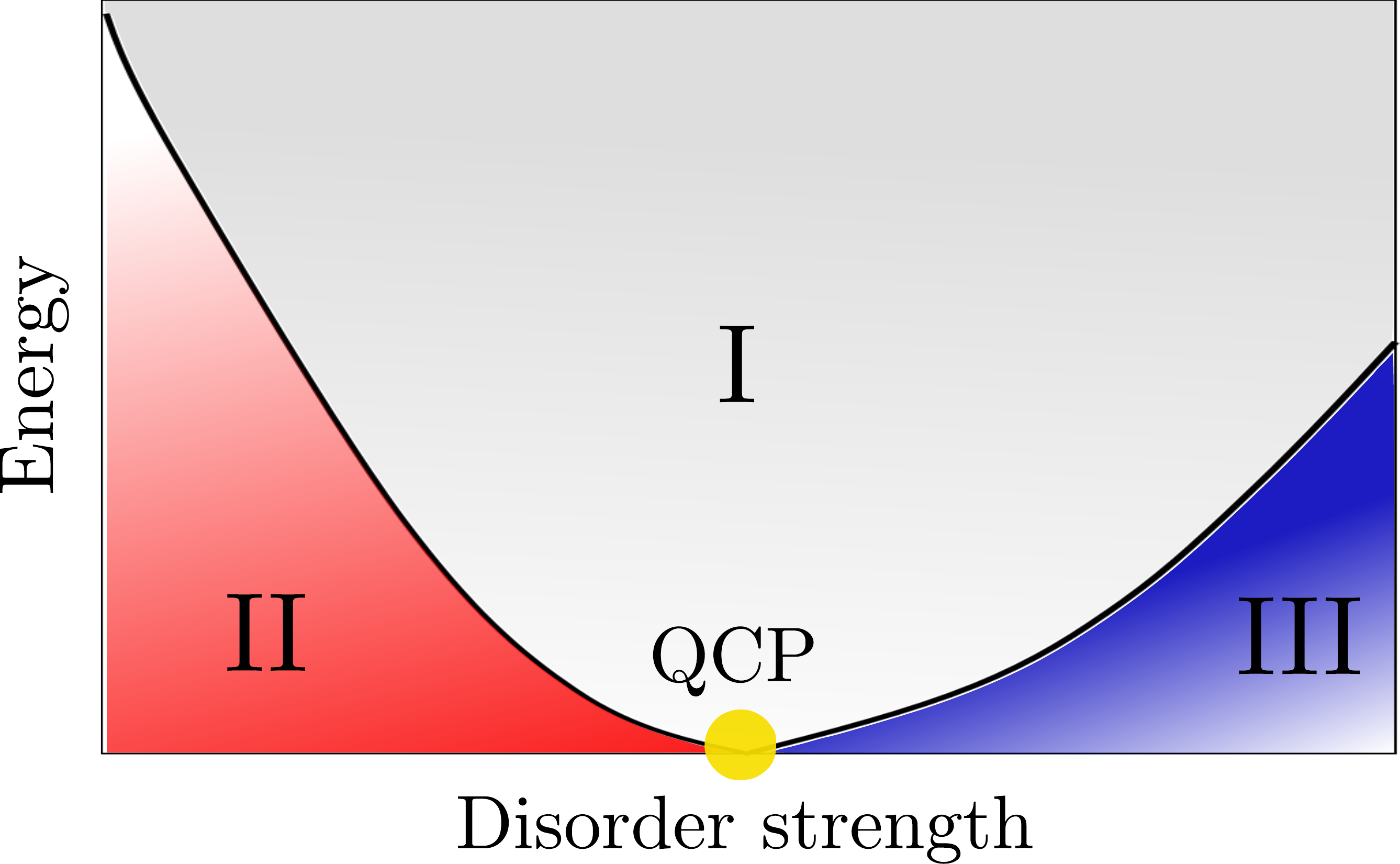}
\centering
\caption{(color online) Schematic energy-disorder phase diagram for disordered Weyl SMs.    }
\label{Fig. 1}
\end{figure}
In Fig.~\ref{Fig. 1}, we present a schematic energy-disorder phase diagram, showing three different regimes distinguished by the energy dependence of the density of states.
In regime I, the energy dependence of the density of states are dominated by the QCP. 
Regime II corresponds to weakly disordered systems at low energy, where the power law is qualitatively equivalent with clean systems.
In regime III, the density of state becomes non-zero at the Weyl point. 
The frequency and temperature dependences of the optical conductivity in these regimes were theoretically predicted in addition to the thermodynamic properties~\cite{roy}.

In this paper, we study the nuclear spin-lattice relaxation rate due to orbital currents in disordered Weyl SMs 
using the self-consistent Born approximation (SCBA) for impurities. 
For the critical exponents $z=2$ and $\nu=1$ within the SCBA, we find that our result is in agreement with the scaling relation of $(T_1T)^{-1}$. 
At $E_\textrm{F}=0$, the nuclear spin-lattice relaxation rate is scaled as $(T_1T)^{-1} \sim T^{2/z}$, $(T_1T)^{-1} \sim T^2$, and $(T_1T)^{-1} \sim (W- W_c)^{2\nu}$ 
in the regimes I, II, and III, respectively, where we denote the disorder strength as $W$ and its critical value as $W_c$.  
We also discuss the relationship to the one-loop renormalization group (RG) analysis, which gives $z=1.5$ \cite{goswami}.
When the Weyl points are away from the Fermi energy, deviations from these behaviors are elucidated with a special emphasis on 
the temperature dependence of $\mu(T)$. 
In particular, we show that a remarkable temperature dependence of $\mu(T)$ caused by impurities amplifies an additional feature of $(T_1T)^{-1}$ 
with a low-temperature upturn in the regime II, which is consistent with the recent NQR experiment on TaP~\cite{yasuoka,okvatovity}. 

The organization of paper is as follows. 
In Section \ref{model}, we introduce the SCBA for a pair of Weyl nodes with the opposite chirality.  In section \ref{transverse}, the impurity averaged transverse conductivity is derived as a function of a wavevector $q$. In order to derive the vertex correction in a gauge-invariant manner, we employ the conserving approximation for the SCBA.
We find that the vertex correction leads to quantitative changes in $(T_1T)^{-1}$ but does not affect the critical behavior.
In section \ref{relaxation_rate}, the nuclear spin-lattice relaxation rate is computed numerically. We also predict the scaling relation of $(T_1T)^{-1}$ from the dimensional analysis, which is in agreement with the SCBA. In section \ref{discussion}, we discuss the effect of the particle-hole asymmetry within the SCTA. In the final section, we provide a summary of this work.

\section{Model Hamiltonian and the disorder-driven quantum criticality}
\label{model}
\subsection{Model}
We consider a pair of Weyl nodes with the opposite chirality. For $2N$ nodes, we simply need to multiply our result by a factor of $N$. With random disorder potential, the Hamiltonian $H=H_0+H_\textrm{D}$ is given as follows. 
\begin{equation}
H_0=\sum_{a=\mathcal{L,R}}\int d\boldsymbol{k}\psi_{k,a}^\dagger \chi_a  \hbar c^* \boldsymbol{k}\cdot\boldsymbol{\sigma}\psi_{k,a},
\end{equation}
\begin{equation}
H_\textrm{D}=\sum_{a,b=\mathcal{L,R}}\int d\boldsymbol{k}\psi_{k,a}^\dagger u_{ab}\psi_{k,b},
\end{equation}
where $c^*$ is the Fermi velocity, $\sigma_i$ is the Pauli matrices and $\mathcal{L/R}$ stands for left/right chiral modes with $\chi_\mathcal{L/R}=\pm1$. The intravalley scattering is assumed to be isotropic ($|u_\mathcal{LL}|^2=|u_\mathcal{RR}|^2$). In addition, the intervalley scattering $u_\mathcal{LR}=u_\mathcal{RL}^*$ is introduced. We should note that a similar model was discussed in graphene~\cite{shon,koshino,arimura,aleiner, ostrovsky}. 

If the range of impurity potential is much shorter than the lattice constant, the intervalley scattering becomes important. In contrast, it is suppressed for the potential with its range comparable to the lattice constant. Following Ref.~\cite{shon}, we consider the short-range scatters ($|u_\mathcal{LR}|^2=|u_\mathcal{LL}|^2$) and the long-range scatters ($|u_\mathcal{LR}|^2=0$). In both of limiting cases, we assume that the range of impurity potential is much smaller than the typical electron wavelength. 

\begin{figure}[t]
\includegraphics[width=80mm]{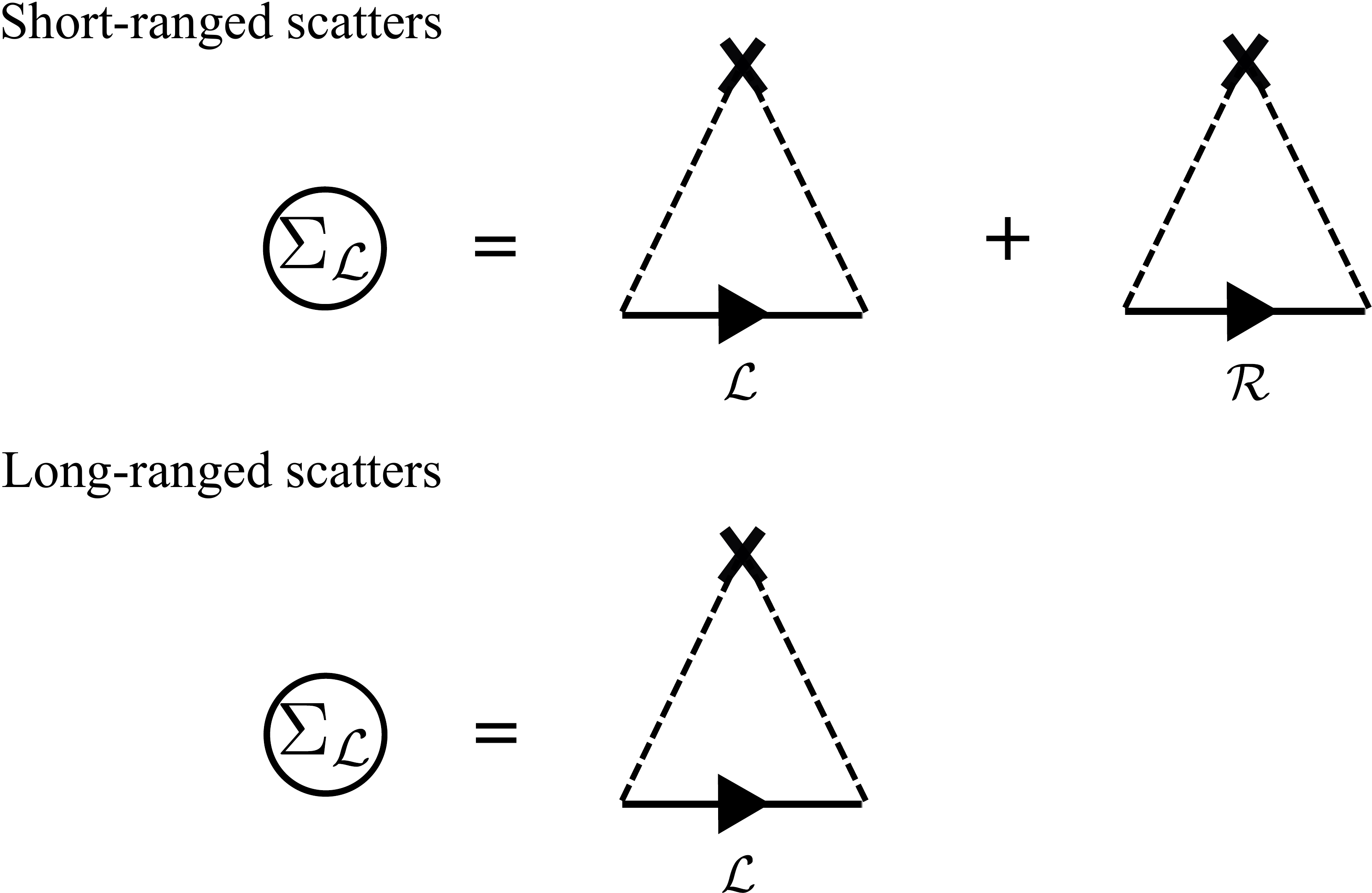}
\centering
\caption{Feynman digram for the self-energy of a Weyl point with left chirality with the short-range scatters ($|u_\mathcal{LR}|^2=|u_\mathcal{LL}|^2$) and the long-range scatters ($|u_\mathcal{LR}|^2=0$). The subscript $\mathcal{L/R}$ implies the chirality.}
\label{Fig. 2}
\end{figure}

In the SCBA, the self-energy is obtained by solving the following self-consistent equation (Figure \ref{Fig. 2}).
\begin{eqnarray}
\Sigma_a^\textrm{R}(\omega)&=&\frac{n_i}{\hbar^2}\sum_{\boldsymbol{k}}\sum_{b=\mathcal{L},\mathcal{R}}|u_{ab}|^2G_b^\textrm{R}(\boldsymbol{k},\omega), 
\label{scba}
\end{eqnarray}
where the the subscript $a$ denotes the chirality $\mathcal{L}/\mathcal{R}$, $n_i$ is the impurity concentration and $G_a^\textrm{R}$ is the retarded Green function after disorder averaging.
The disorder-averaged Green function is defined as 
\begin{equation}
G_a^\textrm{R}(\boldsymbol{k},\omega)=\left[\omega+\mu/\hbar-\chi_a c^*\boldsymbol{\sigma}\cdot\boldsymbol{k}-\Sigma^\textrm{R}_a(\omega)\right]^{-1}.
\label{Green}
\end{equation}
We should note that the self-energy takes the same form for both short-ranged and long-ranged scatters. Assuming $\Sigma^\textrm{R}(\omega)=\Sigma_\textrm{I}^\textrm{R}(\omega)\sigma_0$ with $\sigma_0$ denoting the identity matrix element, the self-consistent equation is simplified to
\begin{eqnarray}
\Sigma_\textrm{I}^\textrm{R}(\omega)&=&\frac{n_i(|u_\mathcal{LL}|^2+|u_\mathcal{RL}|^2)}{\hbar^2}\int\frac{dk^3}{(2\pi)^3}\left[\frac{\tilde{\omega}_\textrm{R}}{(\tilde{\omega}_\textrm{R})^2-c^{*2}k^2}\right]\nonumber\\
&=&W\tilde{\omega}_\textrm{R}f(\omega),
\end{eqnarray}
where $\tilde{\omega}_\textrm{R}=\omega+\mu/\hbar-\Sigma_\textrm{I}^\textrm{R}(\omega)$. 
Since the momentum integration in the above equation is divergent, we introduce the cutoff factor $\frac{k_c^2}{k_c^2+k^2}$ to take account of the finite band width.
The dimensionless impurity strength $W$ and the function $f(\omega)$ are defined as follows.
\begin{align}
W&=\frac{n_iE_c}{2\pi^2\hbar^3 c^{*3}}(|u_\mathcal{LL}|^2+|u_\mathcal{RL}|^2),
\label{imp_st}
\\
f(\omega)&=\int^{\infty}_0 \frac{K^2dK }{K^2+1}\frac{1}{\tilde{\Omega}_\textrm{R}^2-K^2} =-\frac{\pi}{2}\frac{1}{1-i\tilde{\Omega}_\textrm{R}},
\label{func_F}
\end{align}
where the energy cutoff is $E_c=\hbar c^* k_c$ and the dimensionless quantities are $\tilde{\Omega}_\textrm{R}=\hbar\tilde{\omega}_\textrm{R}/E_c$ and $K=\hbar c^*k/E_c$.
The self-consistent solution is obtained as 
\begin{equation}
\tilde{\Omega}_\textrm{R}=\frac{1}{2}\Big(i\delta +\Omega+\textrm{sgn}(\Omega)\sqrt{4i\Omega+(i\delta +\Omega)^2}\Big),
\label{xsol}
\end{equation} 
with $\delta=W/W_c-1$, $W_c=2/\pi$ is the critical impurity strength and $\Omega=(\hbar\omega+\mu)/E_c$. 
The density of states $D(\omega)$ is
\begin{eqnarray}
D(\omega)&=&-\frac{1}{\hbar\pi}\sum_{a=\mathcal{L,R}}\sum_{\boldsymbol{k}}\textrm{Im tr}[G_a^\textrm{R}(\boldsymbol{k},\omega)]\nonumber\\
&=&-\frac{2E_c^2}{(\pi\hbar c^{*})^3}\textrm{Im} \left[\tilde{\Omega}_\textrm{R} f(\omega)\right].
\label{Eq: Dos}
\end{eqnarray}
At $\Omega=0$, we have a simple criterion for the criticality as $\tilde{\Omega}_\textrm{R}=i\delta$ for $\delta>0$ and $\tilde{\Omega}_\textrm{R}=0$ otherwise.

\subsection{Critical exponents in the SCBA}
The disorder-induced quantum criticality is characterized by universal critical exponents. Near the QCP in disordered Weyl SMs, the density of states acts as the order parameter that is described by the critical exponents $z$ and $\nu$. The dynamical exponent $z$ relates the correlation length $\xi$ and the characteristic energy scale $\Omega_0$ as $\Omega_0\sim\xi^{-z}$. At the QCP, the correlation length diverges as $\xi\sim\delta^{-\nu}$.

From the scaling of the density of states, the density of states at the QCP is~\cite{sergev_prb,sergev_rev,kobayashi2}
\begin{equation}
D(\Omega,\delta=0)\sim\Omega^{\frac{d}{z}-1},
\end{equation} 
where $d$ is the spatial dimension of the system.
Above the QCP, the density of states becomes finite even at Weyl nodes. At $\Omega=0$, it is given as
\begin{equation}
D(\Omega=0,\delta>0)\sim\delta^{(d-z)\nu}.
\end{equation}
Below the criticality, the density of states for $\Omega\ll1$ has the same energy dependence as Weyl SMs without disorder.

\begin{figure}[t]
\includegraphics[height=120mm]{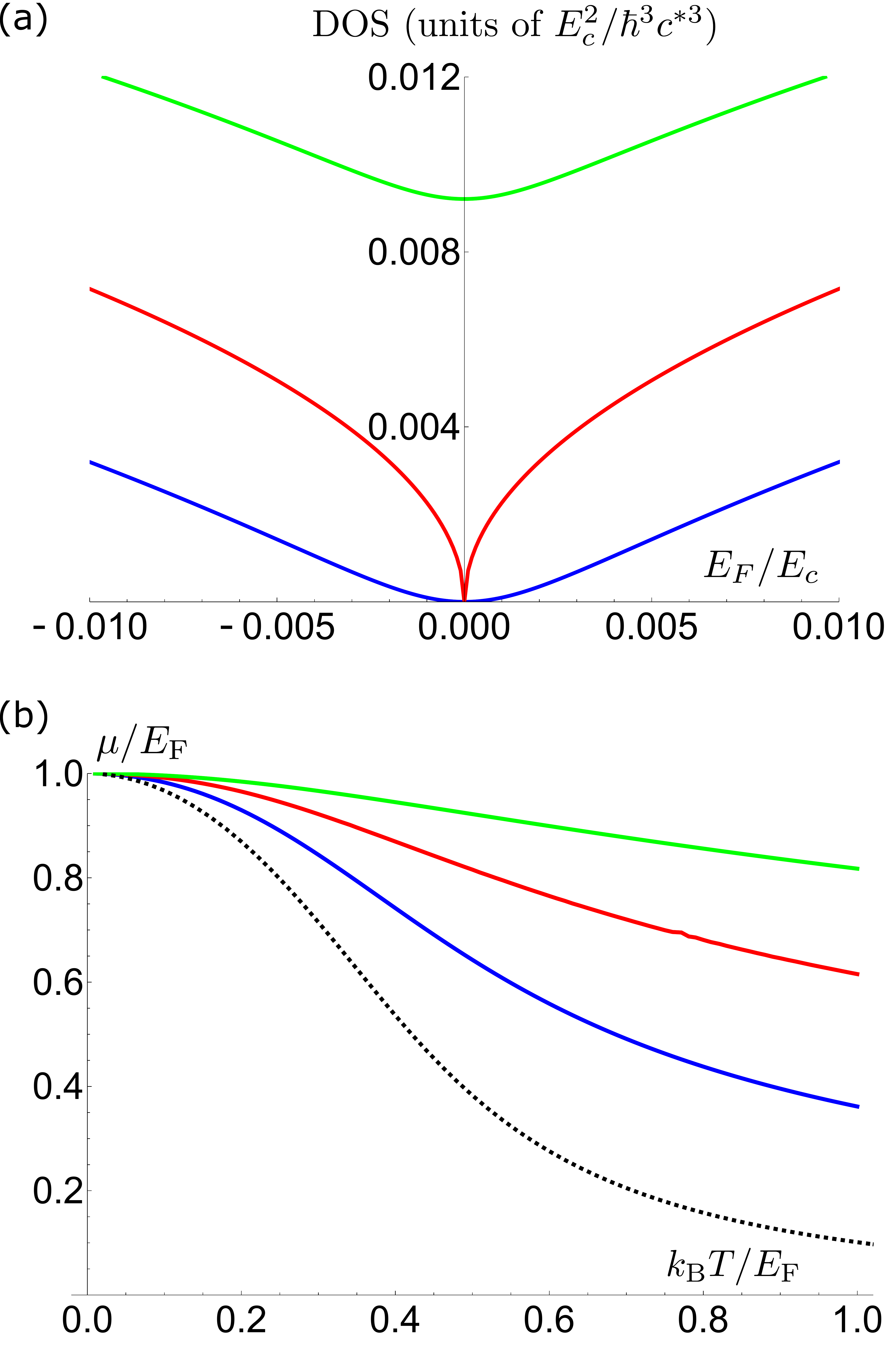}
\centering
\caption{(color online) (a) The density of states in the SCBA against Fermi energy $E_\textrm{F}=\mu(T=0)$. (b) Chemical potential against temperatures for $E_\textrm{F}/E_c=10^{-2}$. A dashed line corresponds to the clean system.  The impurity strength for both plots are set at $W/W_c=0.9 \textrm{ (blue)}, 1.0\textrm{ (red)}$ and $1.1\textrm{ (green)}$ from bottom.
}
\label{fig:dos}
\end{figure}

The solution of Eq.~(\ref{scba}) matches with the result of the saddle-point solution in the limit of $N\rightarrow \infty$ valleys. Thus, the critical exponents in the SCBA are given by $z=2$ and $\nu=1$~\cite{ryu}. 
At the QCP ($\delta= 0$), $\tilde{\Omega}^\textrm{R}\sim (1+i)\,\sqrt{\frac{\Omega}{2}}$ for $\Omega\rightarrow 0$.
The density of states for small $\Omega$ is 
\begin{equation}
D(\Omega)\approx\frac{E_c^2}{\pi^2(\hbar c^{*})^3}\sqrt{\frac{\Omega}{2}},
\label{critical_dos}
\end{equation}
leading to $z=2$. 
In Fig.~\ref{fig:dos}(a), the density of states against Fermi energy $E_\textrm{F}$ is plotted near the critical point $W=W_c$. At the critical point, the gradient of density of states at Weyl points is divergent, showing the root-squared energy dependence. 
Similarly, the density of states at $\Omega=0$ for $\delta>0 $ is given by
\begin{equation}
D(\delta)\approx\frac{ E_c^2}{\pi^2(\hbar c^{*})^3}\delta.
\end{equation}
Hence, $\nu=1$.

\subsection{Chemical potential at finite temperatures}
\label{chemicalsec}
Since the nuclear spin-lattice relaxation rate $1/T_1$ is measured against temperatures, the temperature dependence of chemical potential is important. 
In this section, we consider the effect of impurity on the chemical potential at finite temperatures.
Assuming that a total number of charge carriers is conserved in two bands forming a Weyl cone, we obtain the temperature-dependent chemical potential $\mu(T)$. The particle and hole numbers are given as
\begin{align}
n&=\int^\infty_0dE\,n_\textrm{F}(E)D(E),
\\
p&=\int^0_{-\infty} dE\,(1-n_\textrm{F}(E))D(E).
\end{align}
The change in the total carrier number from $T=0$ to $T=T^\prime$ is given by
\begin{eqnarray}
\delta N&=&(n-p)|_{T=T^\prime}-(n-p)|_{T=0}\nonumber\\
&=&\int^{\infty}_{-\infty}dE\,n_\textrm{F}(E)D(E)-\int^{E_\textrm{F}}_{-\infty}dE\,D(E),
\end{eqnarray}
where $E_\textrm{F}=\mu(T=0)$ is the Fermi energy. In our calculation, chemical potential is numerically computed by keeping $\delta N=0$ with the renormalized density of states under impurity. The integration over energy is taken for a finite width scaled by temperatures, neglecting a small contribution from high energy regions.

In Fig.~\ref{fig:dos}(b), chemical potential is plotted against temperatures near the critical point. In the SCBA, the density of states is symmetric about Weyl nodes.  As a result, chemical potential moves towards Weyl points as the temperature increases. For weak disorder strength, the temperature dependence is almost identical with the clean system, showing a large decrease below $k_\textrm{B} T/E_\textrm{F}\sim 1$. As the impurity strength approaches the critical value, the change in chemical potential becomes smaller.

\section{Transverse Conductivity}
\label{transverse}

\subsection{Formulation}

In this section, we obtain the static conductivity tensor 
$\sigma_{ij}({\boldsymbol{q}})$ for the disordered Weyl fernion systems, 
whose transverse part will be used for computation of $1/T_1$ 
in the next section.
By applying the standard Feynman diagrammatic technique based on the Kubo formula to our systems, the conductivity tensor is given by
\begin{align}
\sigma_{ij}(\boldsymbol{q})
=&
\frac{e^2 {c^*}^2}{\hbar} \textrm{Re}\!\! 
\sum_{|{\boldsymbol{k}}|<k_{\rm c}} \sum_{a=\mathcal{L},\mathcal{R}} 
\int^\infty_{-\infty}\frac{d \omega}{2\pi}
\left(-\frac{\partial n_\textrm{F}(\omega)}{\partial \omega}\right)
\nonumber\\
&\times
\textrm{Tr} \left[\sigma_i G^{\rm R}_a(\boldsymbol{k}_-, \omega)
\Gamma^{{\rm RA}}_{a,j}(\omega\,; \boldsymbol{q}) G^{\rm A}_a(\boldsymbol{k}_+,\omega) 
\right. \nonumber\\ 
& \left. - 
\sigma_i G^{\rm R}_a(\boldsymbol{k}_-,\omega) 
\Gamma^{\rm RR}_{a,j}(\omega\,; \boldsymbol{q}) 
G^{\rm R}_a(\boldsymbol{k}_+,\omega) \right],
\label{conductivity tensor}
\end{align}
where $\boldsymbol{k}_\pm=\boldsymbol{k}\pm\boldsymbol{q}/2$. 
Here the current vertex function is defined as 
$\chi_a c^* \Gamma^{\alpha\beta}_{a,j}(\omega\,; \boldsymbol{q})$ 
with $a=\mathcal{L}/\mathcal{R}$, 
where the superscript $\alpha,\beta$ stands for R/A in relation to 
the retarded/advanced Green functions. 
As in the previous section, we take a smooth cutoff procedure for a wavenumber cutoff $k_{\rm c}$ as
\begin{align}
\sum_{|{\boldsymbol{k}}|<k_{\rm c}} \longrightarrow 
\int \frac{d^3k}{(2\pi)^3} \frac{k_{\rm c}^2}{|{\boldsymbol{k}}|^2 + k_{\rm c}^2}.
\end{align}

\begin{figure}[t]
\includegraphics[width=84mm]{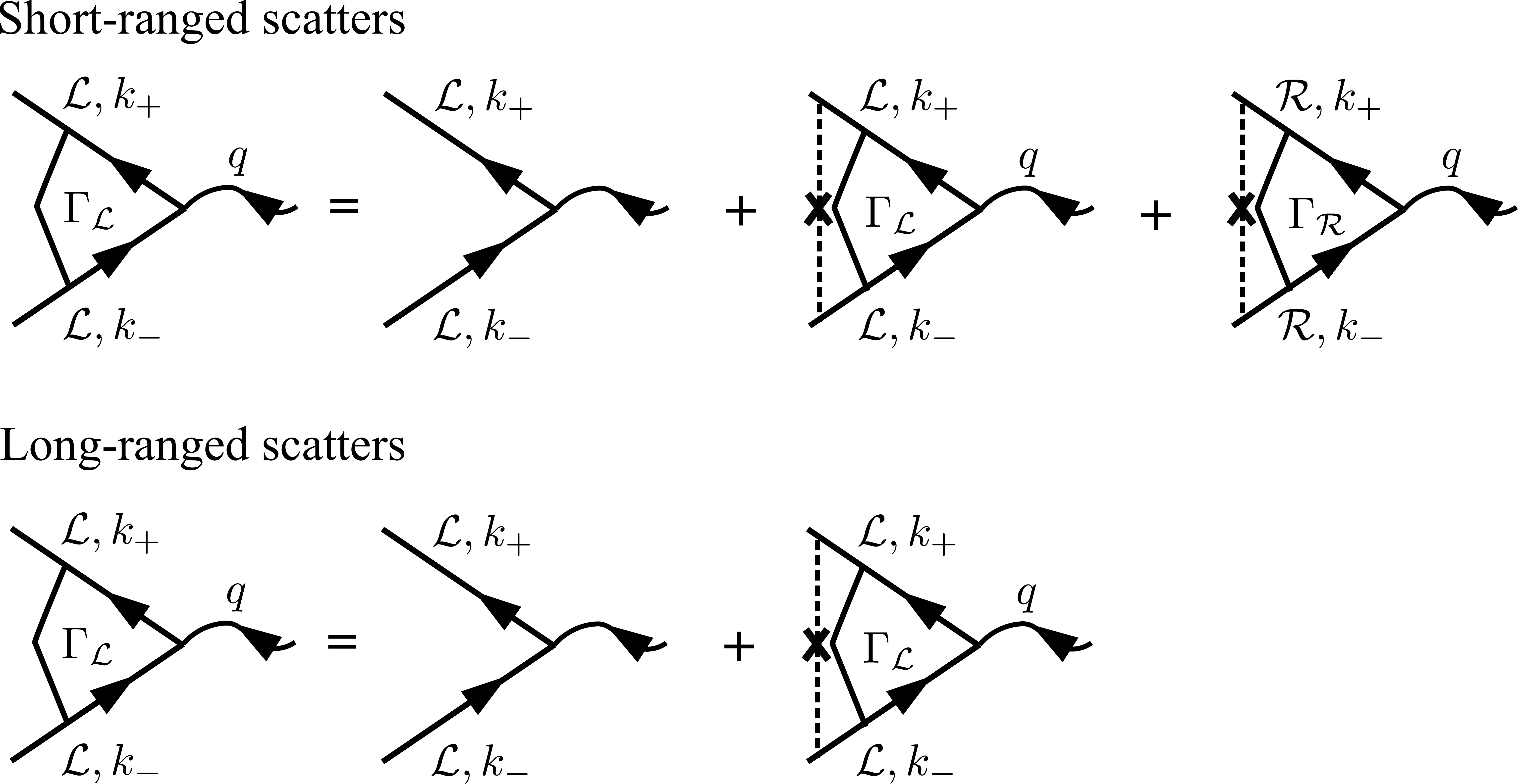}
\centering
\caption{ Feynman digram for the vertex function of a left chiral mode with the short-range scatters ($|u_\mathcal{LR}|^2=|u_\mathcal{LL}|^2$) and the long-range scatters ($|u_\mathcal{LR}|^2=0$). The subscript $\mathcal{L/R}$ implies the chirality. 
}
\label{fig:vertex}
\end{figure}

\subsubsection{Conserving approximation}

To preserve gauge invariance, we introduce a conserving approximation corresponding to the SCBA 
with a special care about the wavenumber cutoff $k_{\rm c}$ of the ${\boldsymbol{k}}$ summation. 
The Bethe-Salpeter equation for the vertex function $\Gamma^{\alpha\beta}_{a,j}(\omega\,; \boldsymbol{q})$
is then given by (Fig.~\ref{fig:vertex})
\begin{align}
\Gamma^{\alpha\beta}_{a,j}(\omega\,; \boldsymbol{q})
=&\, \sigma_j + \frac{n_i}{\hbar^2}\!\!
\sum_{|{\boldsymbol{k}}|<k_{\rm c}}\sum_{b=\mathcal{L},\mathcal{R}}  
\chi_a |u_{ab}|^2 \chi_b
\nonumber \\
&\times 
G^\alpha_b(\boldsymbol{k}_-,\omega) \Gamma_{b,j}^{\alpha\beta}(\omega\,; \boldsymbol{q})
G^\beta_b(\boldsymbol{k}_+,\omega) .
\label{vertex}
\end{align}  
Here the Green functions $G_a^{\alpha} ({\boldsymbol{k}}_\pm,\omega)$ 
in Eqs.~(\ref{conductivity tensor}) and~(\ref{vertex}) 
are redefined so as to incorporate an effect of  the wavenumber cutoff $k_{\rm c}$ as
\begin{align}
G_a^\alpha ({\boldsymbol{k}}_\pm,\omega) \!=\!
\left[ \omega \!+\! \mu/\hbar \!-\! \chi_a c^* 
\boldsymbol{\sigma}\cdot {\boldsymbol{k}}_\pm 
\!-\! \Sigma_a^\alpha (\omega; \pm{\boldsymbol{q}}/2)  \right]^{-1},
\label{ReGreen} 
\end{align}
where the self-consistent equation for the self-energy correction $\Sigma_a^\alpha (\omega; \pm{\boldsymbol{q}}/2)$ 
is given by
\begin{align}
\Sigma_a^\alpha (\omega; \pm{\boldsymbol{q}}/2) 
&=\frac{n_i}{\hbar^2}\sum_{|\boldsymbol{k}|<k_c} \sum_{b=\mathcal{L},\mathcal{R}} 
|u_{ab}|^2 G_b^\alpha ({\boldsymbol{k}}_\pm,\omega).
\label{self-energy correction}
\end{align}

It is, here, emphasized that the self-consistent equation for $\Sigma_a^\alpha (\omega)$ in Sec.~\ref{model} is modified as Eq.~(\ref{self-energy correction}), 
so that the self-energy correction $\Sigma_a^\alpha (\omega; \pm{\boldsymbol{q}}/2)$ in the conductivity tensor becomes dependent on ${\boldsymbol{q}}$. This modification of the self-consistent equation is necessary to preserve gauge invariance for the theory with a finite cutoff $k_{\rm c}$. In fact, by virtue of Eqs.~(\ref{ReGreen}) and~(\ref{self-energy correction}), we find
\begin{align}
&
G^\alpha_a(\boldsymbol{k}_-,\omega)^{-1} - G^\beta_a(\boldsymbol{k}_+,\omega)^{-1}
\nonumber\\
&= \chi_a c^* {\boldsymbol{q}} \cdot \boldsymbol{\sigma}   
+ \frac{n_i }{\hbar^2}\sum_{|\boldsymbol{k}|<k_c} 
\sum_{b=\mathcal{L},\mathcal{R}} |u_{ab}|^2 G_b^\alpha(\boldsymbol{k}_-,\omega)
\nonumber\\
&\quad\times
\left[G_b^\alpha(\boldsymbol{k}_-,\omega)^{-1} -G_b^\beta(\boldsymbol{k}_+,\omega)^{-1} \right]
G_b^\beta(\boldsymbol{k}_+,\omega).
\label{Baym1}
\end{align}
Comparing Eq.~(\ref{Baym1}) with Eq.~(\ref{vertex}), 
we confirm the Ward identity with respect to gauge invariance as
\begin{align}
G^\alpha_a(\boldsymbol{k}_-,\omega)^{-1} - G^\beta_a(\boldsymbol{k}_+,\omega)^{-1}
= \chi_a c^* 
{\boldsymbol{q}} \cdot \boldsymbol{\Gamma}_{a}^{\alpha\beta}(\omega\,; \boldsymbol{q}) .
\label{WT_identity}
\end{align} 
However, we fail to obtain the Ward identity without the modification as in Eq.~(\ref{self-energy correction}).

Now we consider the Bethe-Salpeter equation given by  Eq.~(\ref{vertex}).
Because any $2\times2$ matrix can be written as a linear combination of the Pauli matrices $\sigma_i$ and the identity matrix $\sigma_0 = I_{2\times2}$, 
the vertex function $\Gamma^{\alpha\beta}_{a,j} (\omega;{\boldsymbol{q}})$ is expanded as
$\Gamma^{\alpha\beta}_{a,j} (\omega;{\boldsymbol{q}}) 
= \sum_{\nu = 0}^{3} 
\sigma_\nu \Gamma^{\alpha\beta}_{a,\nu j} (\omega;{\boldsymbol{q}})$, 
where the expansion coefficient is given by
\begin{align}
\Gamma^{\alpha\beta}_{a,\nu j} (\omega;{\boldsymbol{q}})
&= 
\frac{1}{2} {\rm Tr} \left[ \sigma_\nu \Gamma^{\alpha\beta}_{a,j} (\omega;{\boldsymbol{q}}) \right] .
\label{expansion coefficient}
\end{align}
Then we can write Eq.~(\ref{vertex}) as
\begin{align}
\Gamma^{\alpha\beta}_{a,\mu j}(\omega\,; {\boldsymbol{q}})
=&\, \delta_{\mu j} + \frac{n_i E_c}{2\pi^2\hbar^3 c^{*3}}
\sum_{b=\mathcal{L},\mathcal{R}}  |u_{ab}|^2 
\nonumber \\
&\times 
\sum_{\nu = 0}^{3}\Xi^{\alpha\beta}_{b,\mu \nu} (\omega;{\boldsymbol{q}})
\Gamma^{\alpha\beta}_{b,\nu j}(\omega\,; {\boldsymbol{q}}) .
\label{BS2}
\end{align} 
where the dimensionless function 
$\Xi^{\alpha\beta}_{a,\mu \nu} (\omega;{\boldsymbol{q}})$ 
that includes only the self-energy corrections is defined as
\begin{align}
\Xi^{\alpha\beta}_{a,\mu \nu} (\omega;{\boldsymbol{q}})
&=
\frac{\pi^2\hbar c^{*3}}{E_c}
\!\!\sum_{|{\boldsymbol{k}}|<k_{\rm c}}\!\!
{\rm Tr} \left[ 
\sigma_\mu G_a^\alpha(\boldsymbol{k}_-,\omega) \sigma_{\nu} G_a^\beta(\boldsymbol{k}_+,\omega) 
\right] .
\label{ximat}
\end{align}
In the following, we give explicit expressions for the self-energy corrections and the vertex functions to obtain the wavenumber dependent conductivity.

\subsubsection{Self-energy corrections}

To obtain the function $\Xi^{\alpha\beta}_{a,\mu \nu} (\omega;{\boldsymbol{q}})$, 
we assume that the solution of Eq.~(\ref{self-energy correction}) 
is given by
\begin{align}
\Sigma_a^{\alpha} (\omega; \pm{\boldsymbol{q}}/2)
=& \Sigma_\textrm{I}^{\alpha} (\omega; q/2)   \sigma_0
\nonumber
\\
&\pm \frac{\chi_a c^*\boldsymbol{q}\cdot\boldsymbol{\sigma}}{2}
\left[ Z_{\rm L}^{\alpha} (\omega; q/2)  - 1 \right].
\label{SolSCBA'}
\end{align}
Then we can write the Green function, Eq.~(\ref{ReGreen}), as
\begin{align}
G_a^\alpha ({\boldsymbol{k}}_\pm,\omega) 
= \left[\, \tilde{\omega}_\alpha -\chi_a c^*\boldsymbol{\sigma}\cdot( {\boldsymbol{k}}  \pm \tilde{\boldsymbol q}_\alpha/2) \right]^{-1},
\label{Green2}
\end{align}
where $\tilde{\omega}_\alpha=\omega + \mu/\hbar -\Sigma_\textrm{I}^\alpha(\omega; q/2)$ and $\tilde{\boldsymbol q}_\alpha 
=Z_{\rm L}^{\alpha} (\omega; q/2) \,{\boldsymbol{q}}$.

Substituting Eq.~(\ref{Green2}) into Eq.~(\ref{ximat}), 
we find $\Xi^{\alpha\beta}_{a,\mu \nu} (\omega;{\boldsymbol{q}})$ has the form as
\begin{align}
\Xi^{\alpha\beta}_{a,i0}(\omega\,; {\boldsymbol{q}}) 
=&\, \Xi^{\alpha\beta}_{a,0i}(\omega\,; {\boldsymbol{q}}) 
= \chi_a \frac{q_i}{q} {\widetilde \Xi}^{\alpha\beta}_{\rm L} (\omega\,; q) ,
\label{decompose_xi1}
\\
\Xi^{\alpha\beta}_{a,ij}(\omega\,; {\boldsymbol{q}}) 
=&\, \frac{q_iq_j}{q^2}\Xi_{\rm L}^{\alpha\beta} (\omega\,; q) 
+ \left(\delta_{ij}-\frac{q_iq_j}{q^2}\right)\Xi_{\rm T}^{\alpha\beta} (\omega\,; q) 
\nonumber\\
&+ i\chi_a\epsilon_{ijk}\frac{q_k}{q} {\widetilde \Xi}_{\rm T}^{\alpha\beta} (\omega\,; q) ,
\label{decompose_xi2}
\end{align}
where $\epsilon_{ijk}$ is the Levi-Civita tensor.
Here we introduce dimensionless variables ${\tilde \Omega}_\alpha$ and ${\tilde Q}_\alpha$ as
\begin{align}
{\tilde \Omega}_\alpha &= 
\frac{\hbar {\tilde \omega}_\alpha}{E_{\rm c}}
= \frac{\hbar \omega + \mu - \hbar \Sigma_\textrm{I}^\alpha(\omega; q/2)}{E_{\rm c}},
\\
{\tilde Q}_\alpha &= \frac{{\tilde q}_\alpha}{k_{\rm c}}
= Z_{\rm L}^{\alpha} (\omega; q/2) \frac{q}{k_{\rm c}}, 
\end{align}
where $E_{\rm c} = \hbar c^* k_{\rm c}$. 
Then explicit expressions for $\Xi_{\rm L}^{\alpha\beta} (\omega\,; q)$, 
${\widetilde \Xi}_{\rm L}^{\alpha\beta} (\omega\,; q)$, 
$\Xi_{\rm T}^{\alpha\beta} (\omega\,; q)$, and 
${\widetilde \Xi}_{\rm T}^{\alpha\beta} (\omega\,; q)$ 
are obtained as follows:
\begin{align}
\Xi_{\rm L}^{\alpha\beta} (\omega\,; q) =&
\int_0^\infty \!\! \frac{K^2 dK}{1+K^2} \bigg[ \left({\tilde \Omega}_\alpha {\tilde \Omega}_\beta - \frac{1}{4} {\tilde Q}_\alpha {\tilde Q}_\beta - K^2 \right) 
\nonumber\\
& \times I_0^{\alpha\beta} (K) + \frac{1}{2} ({\tilde Q}_\alpha - {\tilde Q}_\beta)I_1^{\alpha\beta}(K)
\nonumber\\
& + 2 I_2^{\alpha\beta}(K) \bigg],
\label{}
\\
{\widetilde \Xi}_{\rm L}^{\alpha\beta} (\omega\,; q) =&
\int_0^\infty \!\! \frac{K^2dK}{1+K^2}  \left[ \frac{1}{2}  
\left({\tilde \Omega}_\alpha {\tilde Q}_\beta  - 
{\tilde \Omega}_\beta {\tilde Q}_\alpha \right) I_0^{\alpha\beta}(K)
\right. 
\nonumber\\
& \left. + ({\tilde \Omega}_\alpha + {\tilde \Omega}_\beta)I_1^{\alpha\beta}(K) \right],
\label{}
\\
\Xi_{\rm T}^{\alpha\beta} (\omega\,; q) =&
\int_0^\infty \!\! \frac{K^2dK}{1+K^2}  \left[ \left({\tilde \Omega}_\alpha {\tilde \Omega}_\beta +\frac{1}{4} {\tilde Q}_\alpha {\tilde Q}_\beta \right) 
I_0^{\alpha\beta}(K) \right. 
\nonumber\\
& \left. - \frac{1}{2} ({\tilde Q}_\alpha - {\tilde Q}_\beta)I_1^{\alpha\beta}(K)
- I_2^{\alpha\beta}(K) \right],
\label{transverse1}
\\
{\widetilde \Xi}_{\rm T}^{\alpha\beta} (\omega\,; q) =&
\int_0^\infty \!\! \frac{K^2dK}{1+K^2}  \left[ \frac{1}{2}  
\left({\tilde \Omega}_\alpha {\tilde Q}_\beta  + 
{\tilde \Omega}_\beta {\tilde Q}_\alpha \right)
I_0^{\alpha\beta}(K) \right. 
\nonumber\\
& \left. + ({\tilde \Omega}_\alpha - {\tilde \Omega}_\beta)I_1^{\alpha\beta}(K) \right],
\label{transverse2}
\end{align}
where the integrals $I_n^{\alpha\beta}(K)$ for $n = 0,1,2$ are given by
\begin{align}
I_n^{\alpha\beta}(K) =&\,K^{n} \!\! \int_{-1}^{1} \frac{d x}{2} 
x^n \!
\left( {\tilde \Omega}_\alpha^2 \!-\! \frac{1}{4} {\tilde Q}_\alpha^2 \!-\! K^2 \!+\! K{\tilde Q}_\alpha x \right)^{-1}
\nonumber\\
&\times \left( {\tilde \Omega}_\beta^2 \!-\! \frac{1}{4} {\tilde Q}_\beta^2 \!-\! K^2 \!-\! K{\tilde Q}_\beta x \right)^{-1} .
\end{align}
In particular, 
${\widetilde \Xi}_{\rm L}^{\rm RR} (\omega\,; q) = 0$ 
because of $I_1^{\rm RR}(K)=0$.

\subsubsection{Vertex functions}
Here, we obtain the expressions for vertex functions, which do not vanish for $q>0$ in general.
The solution of Eq.~(\ref{BS2}) can be obtained in the same form 
as Eqs.~(\ref{decompose_xi1}) and~(\ref{decompose_xi2}) to find
\begin{align}
\Gamma^{\alpha\beta}_{a,i0}(\omega\,; {\boldsymbol{q}}) 
=&\, \Gamma^{\alpha\beta}_{a,0i}(\omega\,; {\boldsymbol{q}}) 
= \chi_a \frac{q_i}{q} 
{\widetilde \Gamma}^{\alpha\beta}_{\rm L} (\omega\,; q) ,
\label{decompose_Gamma1}
\\
\Gamma^{\alpha\beta}_{a,ij}(\omega\,; {\boldsymbol{q}}) 
=&\, \frac{q_iq_j}{q^2} \Gamma_{\rm L}^{\alpha\beta} (\omega\,; q) 
+ \left(\delta_{ij}-\frac{q_iq_j}{q^2}\right) \Gamma_{\rm T}^{\alpha\beta} (\omega\,; q) 
\nonumber\\
&+ i\chi_a\epsilon_{ijk}\frac{q_k}{q} {\widetilde \Gamma}_{\rm T}^{\alpha\beta} (\omega\,; q) .
\label{decompose_Gamma2}
\end{align}
By virtue of the Ward identity, Eq.~(\ref{WT_identity}), 
the longitudinal vertex functions 
$\Gamma_{\rm L}^{\alpha\beta} (\omega\,; q)$ and 
${\widetilde \Gamma}^{\alpha\beta}_{\rm L} (\omega\,; q)$ 
are directly related to the self-energy correction, 
Eq.~(\ref{SolSCBA'}), as 
\begin{align}
\Gamma_{\rm L}^{\rm RR} (\omega\,; q)  
&= Z_{\rm L}^{\rm R} (\omega;q/2) ,
\\
{\widetilde \Gamma}_{\rm L}^{\rm RR} (\omega\,; q)  
&= 0 ,
\\
\Gamma_{\rm L}^{\rm RA} (\omega\,; q)  
&= {\rm Re}  Z_{\rm L}^{\rm R} (\omega;q/2)  ,
\\
{\widetilde \Gamma}_{\rm L}^{\rm RA} (\omega\,; q)  
&= - \frac{2 i}{c^*q} {\rm Im} \Sigma_\textrm{I}^{\rm R} (\omega; q/2)   .
\end{align}
On the other hand, 
the transverse vertex functions $\Gamma_{\rm T}^{\alpha\beta} (\omega\,; q)$ 
and ${\widetilde \Gamma}_{\rm T}^{\alpha\beta} (\omega\,; q)$ cannot be determined from the Ward identity. 
They are  given in terms of $\Xi_{\rm T}^{\alpha\beta}(\omega\,; q)$ and ${\widetilde \Xi}_{\rm T}^{\alpha\beta}(\omega\,; q)$ as
\begin{align}
\Gamma_{\rm T}^{\alpha\beta} (\omega\,; q) \!=&
[ 1 \!-\! W \Xi_{\rm T}^{\alpha\beta}(\omega\,; q) ]
\Big/
\Big[ [1 \!-\! W \Xi_{\rm T}^{\alpha\beta}(\omega\,; q)]
\nonumber\\
&\times [1 \!-\! W_- \Xi_{\rm T}^{\alpha\beta}(\omega\,; q)]
\!-\! WW_- {\widetilde \Xi}_{\rm T}^{\alpha\beta}(\omega\,; q)^2
\Big],
\label{GammaT1}
\\
{\widetilde \Gamma}_{\rm T}^{\alpha\beta}  (\omega\,; q) \!=&
W {\widetilde \Xi}_{\rm T}^{\alpha\beta}(\omega\,; q)
\Big/\Big[
[1 \!-\! W \Xi_{\rm T}^{\alpha\beta}(\omega\,; q)]
\nonumber\\
&\times [1 \!-\! W_- \Xi_{\rm T}^{\alpha\beta}(\omega\,; q)]
\!-\! WW_- {\widetilde \Xi}_{\rm T}^{\alpha\beta}(\omega\,; q)^2
\Big],
\label{GammaT2}
\end{align}
where $W_{-}$ is given by
\begin{align}
W_{-} = \frac{n_i E_c}{2\pi^2\hbar^3 c^{*3}} 
\left( |u_{\mathcal{L}\mathcal{L}}|^2 - |u_{\mathcal{L}\mathcal{R}}|^2 \right).
\end{align}

\subsubsection{Wavenumber dependent conductivity}

By using Eqs.~(\ref{expansion coefficient}) and~(\ref{ximat}), 
we write the conductivity tensor, Eq.~(\ref{conductivity tensor}), as 
\begin{align}
\sigma_{ij}(\boldsymbol{q})
=&\, \frac{e^2E_c}{2 \pi^3\hbar^2 c^{*}}\!  
\int^\infty_{-\infty} d \omega
\left(-\frac{\partial n_\textrm{F}(\omega)}{\partial \omega}\right)
\nonumber 
\\
&\times \sum_{a=\mathcal{L},\mathcal{R}}  \sum_{\mu = 0}^{3}
\textrm{Re} \left[\Xi^{{\rm RA}}_{a,i \mu} (\omega;{\boldsymbol{q}})
\Gamma^{{\rm RA}}_{a,\mu j}(\omega\,; {\boldsymbol{q}}) 
\right. \nonumber\\ &\left. 
\hspace{2.2cm} 
- \Xi^{{\rm RR}}_{a,i \mu} (\omega;{\boldsymbol{q}})
\Gamma^{{\rm RR}}_{a,\mu j}(\omega\,; {\boldsymbol{q}})  
 \right],
\label{}
\end{align}
Substituting Eqs.~(\ref{decompose_xi1}), (\ref{decompose_xi2}), (\ref{decompose_Gamma1}), and (\ref{decompose_Gamma2})
into this equation, we find the conductivity tensor has the form as
\begin{align}
\sigma_{ij}({\boldsymbol{q}}) = \frac{q_iq_j}{q^2}\sigma_{\rm L}(q)
+ \left(\delta_{ij} - \frac{q_iq_j}{q^2} \right) \sigma_{\rm T}(q).
\label{}
\end{align}
It is to be noted that the term including $\epsilon_{ijk}$ 
similar to the third terms in Eqs.~(\ref{decompose_xi2}) and~(\ref{decompose_Gamma2}) vanishes by taking the summation with respect to the chirality $a=\mathcal{L}/\mathcal{R}$.
Then we obtain the transverse conductivity $\sigma_{\rm T}(q)$ as
\begin{align}
\sigma_{\rm T}(q)
=&\, \sigma_\textrm{unit}\!
\int^\infty_{-\infty}d\omega \! \left(-\frac{\partial n_{\rm F}(\omega)}{\partial \omega}\right)\!
\textrm{Re}\! \left[ 
\Xi_{\rm T}^{\rm RA} (\omega\,; q) \Gamma_{\rm T}^{\rm RA} (\omega\,; q)
\right.
\nonumber\\
&
\,\,
- \Xi_{\rm T}^{\rm RR} (\omega\,; q) \Gamma_{\rm T}^{\rm RR} (\omega\,; q)
+ {\widetilde \Xi}_{\rm T}^{\rm RA} (\omega\,; q) {\widetilde \Gamma}_{\rm T}^{\rm RA} (\omega\,; q)
\nonumber\\
&
\,\, \left.
- \,{\widetilde \Xi}_{\rm T}^{\rm RR} (\omega\,; q) {\widetilde \Gamma}_{\rm T}^{\rm RR} (\omega\,; q)
\right] ,
\label{sigmaq_full}
\end{align}
and the longitudinal conductivity $\sigma_{\rm L}(q)$ 
by replacing the subscript ${\rm T}$ by ${\rm L}$ in Eq.~(\ref{sigmaq_full}).
Here $\sigma_\textrm{unit}$ has the dimension of conductivity as
\begin{equation}
\sigma_\textrm{unit} = \frac{e^2E_c}{\pi^3\hbar^2 c^{*}} .
\end{equation}

\subsection{Results}

As derived in Appendix \ref{approximate solution}, an approximate solution of Eq.~(\ref{self-energy correction}) for $q/k_{c} \ll 1$ are given by Eq.~(\ref{SolSCBA'}) with
\begin{align}
\tilde{\Omega}_\textrm{R} (\omega)&=
\frac{1}{2}\Big(i\delta +\Omega+\textrm{sgn}(\Omega)\sqrt{4i\Omega+(i\delta +\Omega)^2}\Big),
\label{sol_1}
\\
Z_{\rm L}^{\rm R}(\omega) &= 
1 -\frac{W_-\pi(i+2\tilde{\Omega}_\textrm{R} (\omega)) }{\pi W_-[i+2\tilde{\Omega}_\textrm{R} (\omega)]-12i [i+\tilde{\Omega}_\textrm{R} (\omega)]^2},
\label{sol_2}
\end{align}
where $\delta=W/W_c-1$ and $\Omega=(\hbar\omega+\mu)/E_c$.
Since we are interested in the $q$ dependence of the transverse conductivity $\sigma_{\rm T}(q)$ for $q/k_{c} \ll 1$, we take Eqs.~(\ref{sol_1}) and~(\ref{sol_2}) as the self-energy corrections.
Here, ${\tilde \omega}_{\rm R}$ in Eq.~(\ref{Green2}) is the same as in the previous section. However, there is an important correction in $\tilde{\boldsymbol q}_{\rm R}$ for the long-range scatters with $W_- \neq 0$.
To elucidate the difference between short-range and long-range scatters, we perform an explicit calculation of
$\Xi_{\rm T}^{\alpha\beta} (\omega\,; q)$ and ${\widetilde \Xi}_{\rm T}^{\alpha\beta} (\omega\,; q)$ 
given by Eqs.~(\ref{transverse1}) and~(\ref{transverse2}), respectively.
The transverse conductivity $\sigma_{\rm T}(q)$ is obtained from evaluating Eq.~(\ref{sigmaq_full}) together with Eqs.~(\ref{GammaT1}) and~(\ref{GammaT2}).

\begin{figure}[t]
\includegraphics[width=78mm]{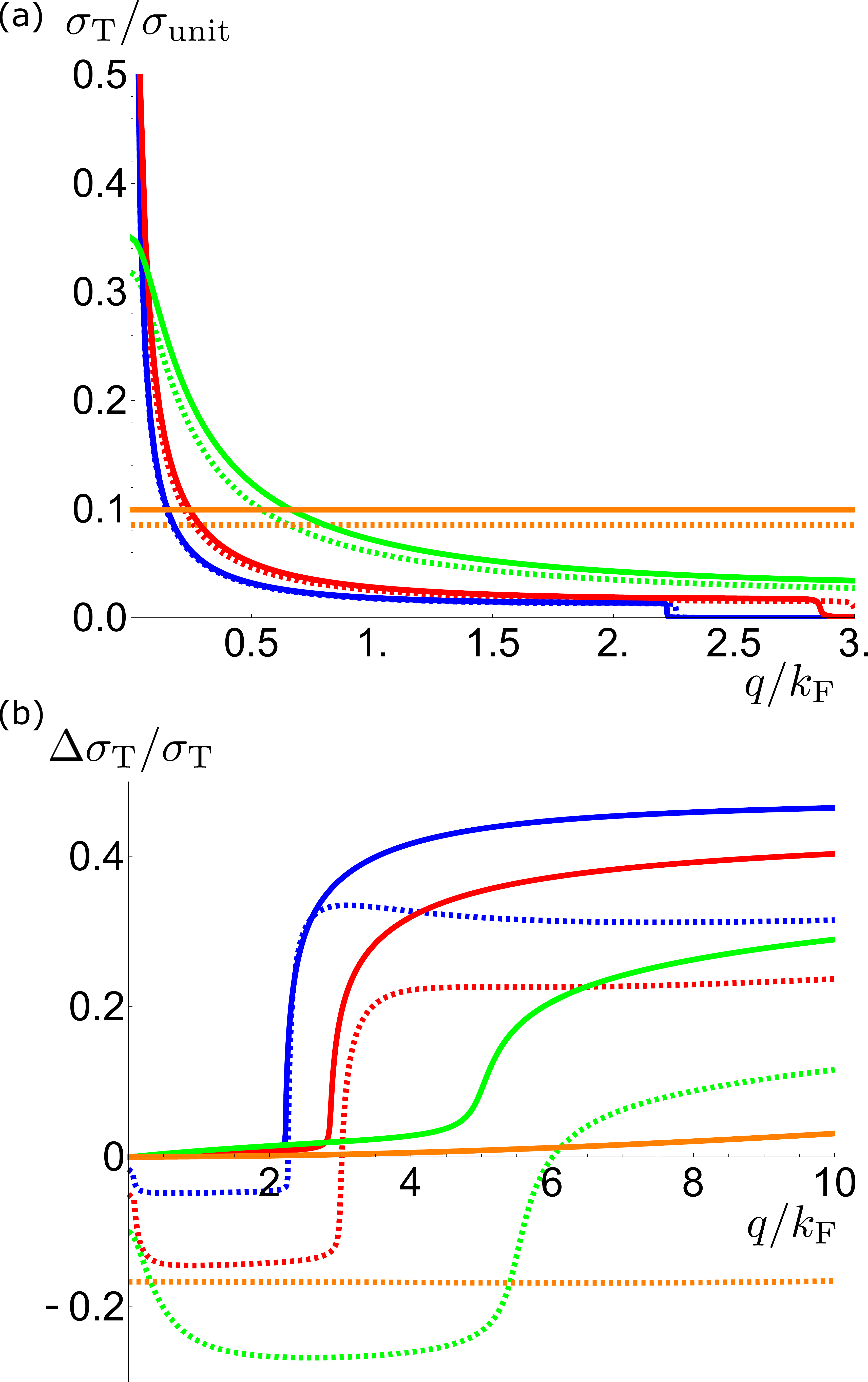}
\centering
\caption{(color online) Plot of (a) the transverse conductivity $\sigma_\textrm{T}(q)$ and (b) the ratio $\Delta\sigma_\textrm{T}/\sigma_\textrm{T}$ at $T=0$ and $E_\textrm{F}/E_c=10^{-2}$ for short-range scatters (solid lines) and long-ranged scatters (dashed lines). 
The disorder strength is taken at $W/W_c=0.1$ (blue), $0.3$ (red), $0.6$ (green) and $1.0$ (orange) from (a) bottom and (b) top.  }
\label{sigmaq}
\end{figure}

\begin{figure}
  \includegraphics[width=75mm]{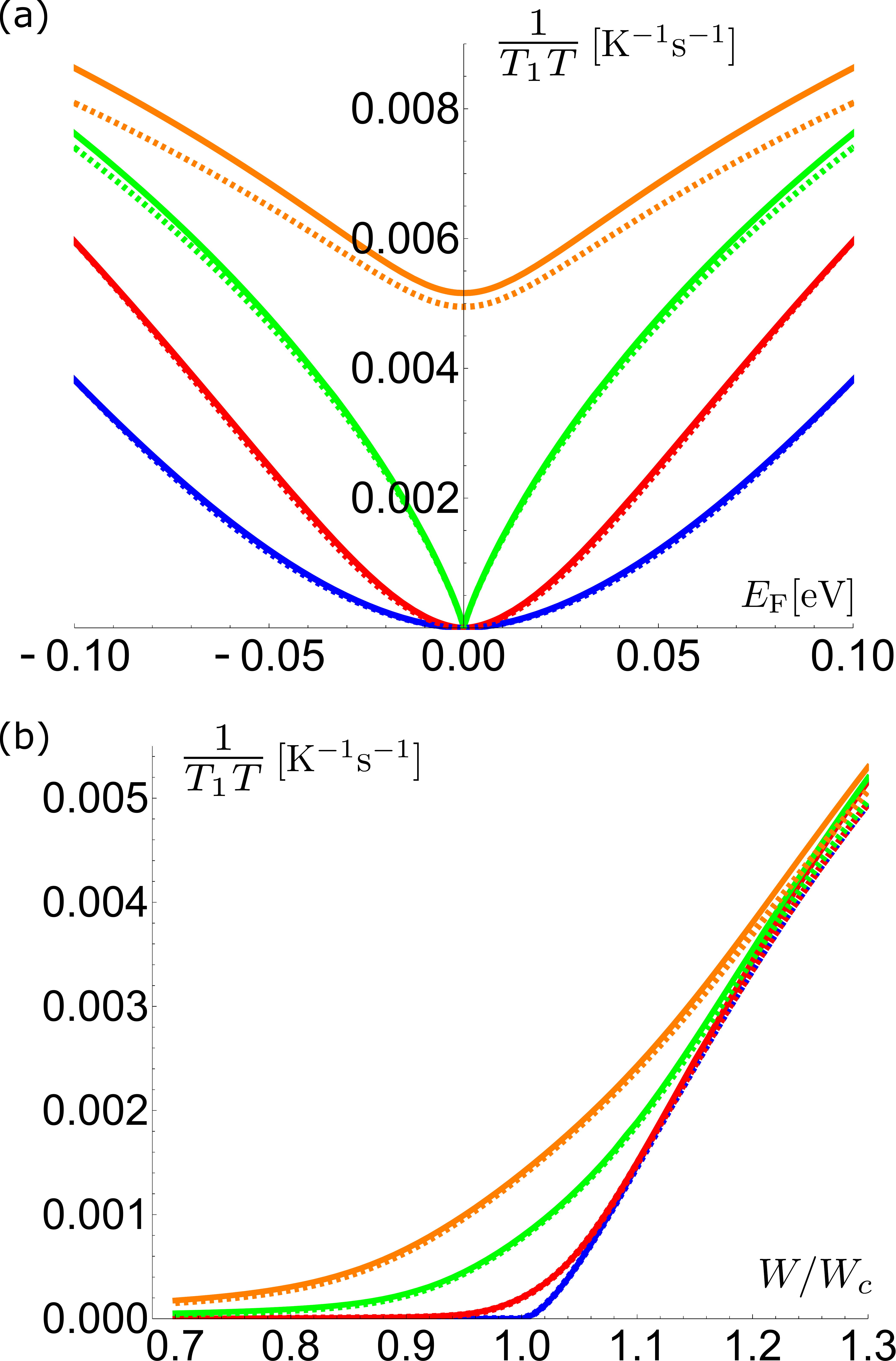}
  \caption{ (color online) The nuclear spin-lattice relaxation rate $(T_1T)^{-1}$ at $T=0$ (a) against the Fermi energy and (b) against the disorder strength for short-range scatters (solid lines) and long-ranged scatters (dashed lines). In (a), the disorder strength is taken at $W/W_c=0.4$ (blue), $0.7$ (red), $1.0$ (green) and $1.3$ (orange) from bottom. In (b), the Fermi energy is taken at $E_\textrm{F}=0$ (blue), $1$ meV (red), $5$ meV (green) and $10$ meV (orange) from bottom. 
 }
\label{rate1}
\end{figure}

Figure \ref{sigmaq}(a) shows the $q$ dependence of transverse conductivity at $T=0$ K and $E_\textrm{F}/E_c=10^{-2}$ for short-range and long-range scatters.
The behavior of $\sigma_\textrm{T}(q)$ is essentially identical for both cases at various disorder strengths. 
In the weak disorder regime ($W/W_c=0.1$), it is peaked at $q=0$ and converges to zero near $q=2k_\textrm{F}$. 
The long-ranged scatters shows a longer tail at $q>2k_\textrm{F}$ than the short-ranged scatters due to the shift in $q$ by $Z_{\rm L}^{\rm R}(\omega) $.
As the impurity strength is increased, the sharp peak at $q=0$ is broadened and it acquires a longer tail for $q>2k_\textrm{F}$. At the QCP ($W/W_c=1.0$), the transverse conductivity becomes constant for $q\ll k_c$. 
In this regime, the negative contribution of the vertex correction results in the smaller transverse conductivity for the long-ranged scatters at $q\sim k_\textrm{F}$.

Secondly, we consider the $q$ dependence of the vertex correction. 
The transverse conductivity is separated into the contribution from the bare vertex ($\Gamma_i=\sigma_i$) and the full vertex function 
\begin{equation}
\sigma_\textrm{T}(q)=\sigma_\textrm{T}^{(0)}(q)+\Delta\sigma_\textrm{T}(q),
\end{equation}
where the transverse conductivity with the bare vertex is given by
\begin{align}
\sigma_{\rm T}^{(0)}(q)
=&\, \sigma_\textrm{unit}\!
\int^\infty_{-\infty}d\omega \! \left(-\frac{\partial n_{\rm F}(\omega)}{\partial \omega}\right)\!
\nonumber\\
&\times\textrm{Re}\, [ \Xi_{\rm T}^{\rm RA} (\omega\,; q) - \Xi_{\rm T}^{\rm RR} (\omega\,; q) ].
\label{nov_cond}
\end{align}
In Fig.~\ref{sigmaq}(b), the ratio $\Delta\sigma_\textrm{T}/\sigma_\textrm{T}$ is plotted for short-range scatters and long-range scatters, respectively. 
For the short-ranged scatters, the vertex correction vanishes at $q=0$. This is because $\Gamma_\textrm{T}^{\textrm{RA/RR}}(q)=1$ for the short-ranged scatters and ${\widetilde \Xi}_{\rm T}^{\textrm{RA/RR}} (\omega\,; q=0)=0$. Also, the effect of positive contributions for $q>2k_\textrm{F}$ is limited as the transverse conductivity is vanishingly small.  
At the QCP, the vertex correction is negligible for $q\ll k_c$.
In contrast, the vertex correction is important for the long-ranged scatters, showing negative contributions as the disorder strength is increased. 
At the QCP, the vertex correction accounts for approximately $-20\%$ of $\sigma_\textrm{T}$ for $q\ll k_c$.

\section{Nuclear spin-lattice relaxation rate}
\label{relaxation_rate}

As mentioned in Sec. \ref{Introduction}, 
the nuclear spin-lattice relaxation rate $1/T_1$ due to orbital currents  is generally related to the real part of the dynamical transverse conductivity $\sigma_{\rm T} (q, \omega_0)$ with the nuclear Larmor frequency $\omega_0$. 
It is, however, noted that the present systems with disorder have a 
finite damping rate $1/\tau(\omega) = -2 {\rm Im}[ \Sigma^{\rm R}(\omega)]$. For $\omega_0 \tau(\omega_0) \ll 1$, 
the dynamical conductivity can be approximated by the static conductivity as ${\rm Re} \,\sigma_{\rm T} (q, \omega_0) \approx \sigma_{\rm T} (q)$.
From Eq.~(\ref{Eq. 1}), the nuclear spin-lattice relaxation rate is given by
\begin{eqnarray}
\frac{\hbar}{T_1k_\textrm{B}T}&=&\frac{2\gamma_\textrm{n}^2\mu_0^2\hbar}{3 \pi^2}\int^\infty_0 dq\,\sigma_{\rm T}(q) \,
\frac{q_{\rm c}^2}{q^2 + q_{\rm c}^2},
\label{numerical_rate}
\end{eqnarray}
where we introduce a smooth cutoff procedure with 
$q_{\rm c} = 2 k_{\rm c}$. 
By substituting Eq.~(\ref{numerical_rate}) for $\sigma_\textrm{T}(q)$, 
we compute $(T_1T)^{-1}$ for the disordered Weyl fernion systems 
and elucidate its critical behavior near the QCP.

\subsection{Numerical Results}
\label{sec: numerical_rate}
In the previous section, we have shown that the difference between the short-ranged and long-ranged scatters on $\sigma_\textrm{T}(q)$ is resulted from the self-energy and the vertex correction.
In order to confirm that it does not affect the critical behavior, we compare the nuclear spin-lattice relaxation rate for both cases at $T=0$. 
In the following, the parameters are fixed as $E_c=1.0$ eV, $c^*=10^4$ m/s, and $\gamma_\textrm{n}=267.5\,\textrm{s}^{-1}\textrm{T}^{-1}$.

In Fig.~\ref{rate1}(a),  the nuclear spin-lattice relaxation rate is plotted as a function of the Fermi energy at $T=0$. 
While there is no significant difference in $(T_1T)^{-1}$ between the short-ranged and long-ranged scatters below the QCP, their difference becomes visible above the QCP.
At the QCP, $(T_1T)^{-1}$ is proportional to $E_\textrm{F}$ for both scatters for small $E_\textrm{F}$. 
The small difference between the short-ranged and long-ranged scatters is resulted from the cancellation of  the negative vertex correction by the self-energy correction $Z_{\rm L}^{\rm R}(\omega)$ after performing the integration.
In addition to the energy dependence, the nuclear spin-lattice relaxation rate scales with the disorder strength, as shown in Fig.~\ref{rate1}(b). 
We find that $(T_1T)^{-1}$ is proportional to $(W-W_c)^2$ in the limit of $E_\textrm{F}=T=0$, although it deviates from the quadratic dependence away from the QCP.

\begin{figure}
  \includegraphics[width=75mm]{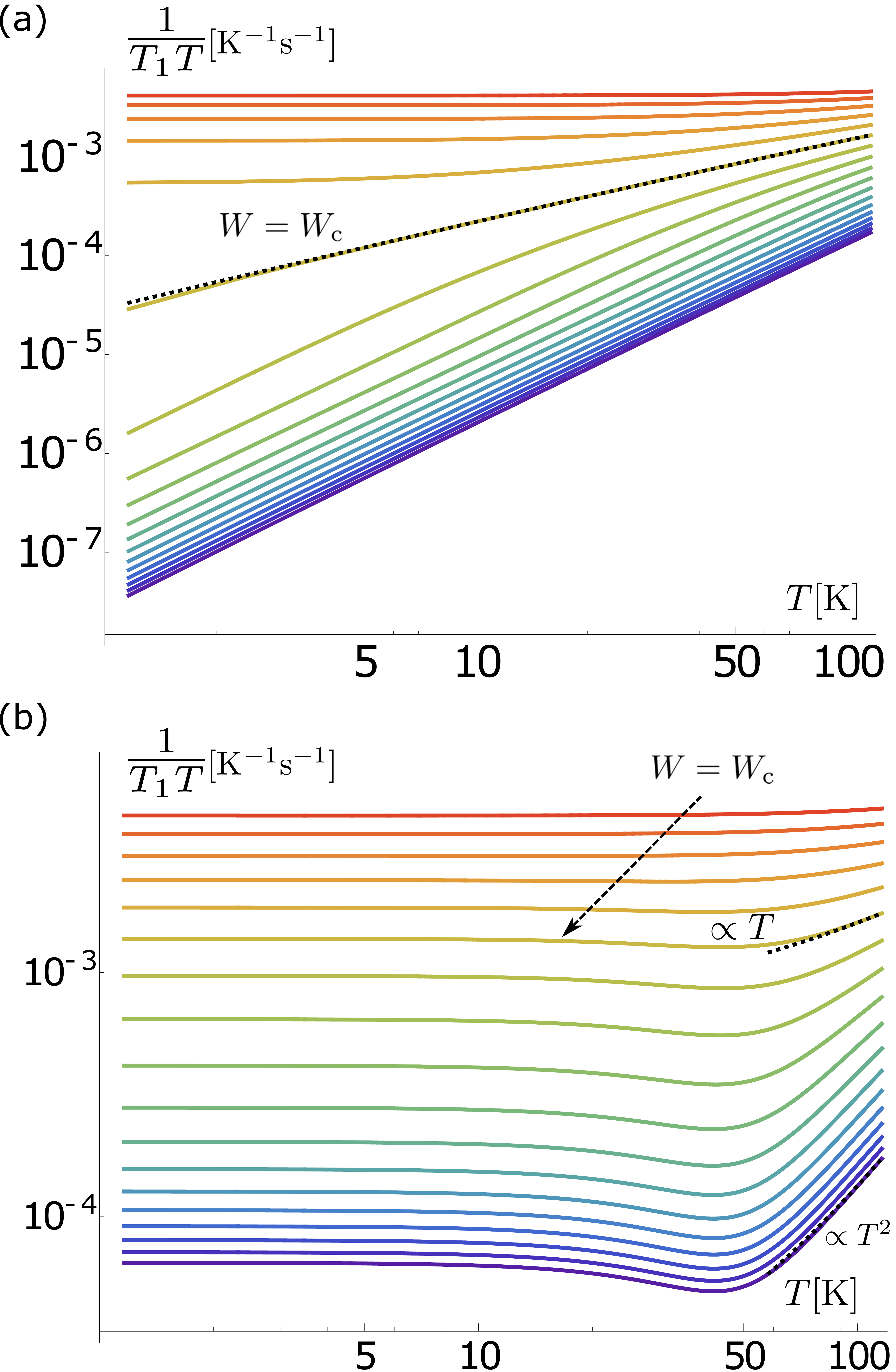}
  \caption{ (color online) The nuclear spin-lattice relation rate $(T_1T)^{-1}$ against temperatures for long-ranged scatters (a) at $E_\textrm{F}=0$ and (b) $E_\textrm{F}=10$ meV, respectively.
The impurity strength $W$ is taken between $0.4$ and $1.3$ with increase by $0.05$. In (a), the dashed line represents the asymptotic expression Eq.~(\ref{critical_rate_mu0}) with $C_1=4.0$ and $C_2=3.8$.  
 }
\label{rate2}
\end{figure}

Since the critical behavior is the same for both the short-ranged and long-ranged scatters, we consider the nuclear spin-lattice relaxation rate under the long-ranged scatters at finite temperatures.
Firstly, the nuclear spin-lattice relaxation rate at $E_\textrm{F}=0$ is considered. In this case, the chemical potential does not depend on the temperature. In Fig.~\ref{rate2}(a), we plot $(T_1T)^{-1}$ as a function of temperatures for different impurity strengths, which shows the critical behavior of the nuclear spin-lattice relaxation rate.
As the impurity strength increases, the transition from $(T_1T)^{-1}\propto T^2$ to $T$ occurs at lower temperatures.
At the QCP,  the nuclear spin-lattice relaxation rate is linearly proportional to the temperature from $T=0$.
Above the QCP, the density of states at the Weyl point becomes finite so $(T_1T)^{-1}$ becomes roughly constant.

Secondly, we consider the nuclear spin-lattice relaxation rate for $E_\textrm{F}\ne 0$.
In Fig.~\ref{rate2}(b), we present the temperature dependence of $(T_1T)^{-1}$ at $E_\textrm{F}=10$ meV.
At low temperatures, $(T_1T)^{-1}$ is constant regardless of the impurity strength.
For $k_\textrm{B}T\sim \mu(T)\sim E_\textrm{F}/2$, $(T_1T)^{-1}$ initially shows a decrease followed by an increase with the strong temperature dependence.
This upturn in $(T_1T)^{-1}$ is caused by the shift in chemical potential~\cite{okvatovity}.   
Above $k_\textrm{B}T\sim E_\textrm{F}/2$, it shows the transition from the $T^2$ dependence for weakly disordered systems to the $T$-linear behavior at the QCP.

\begin{figure}[t]
\includegraphics[width=78mm]{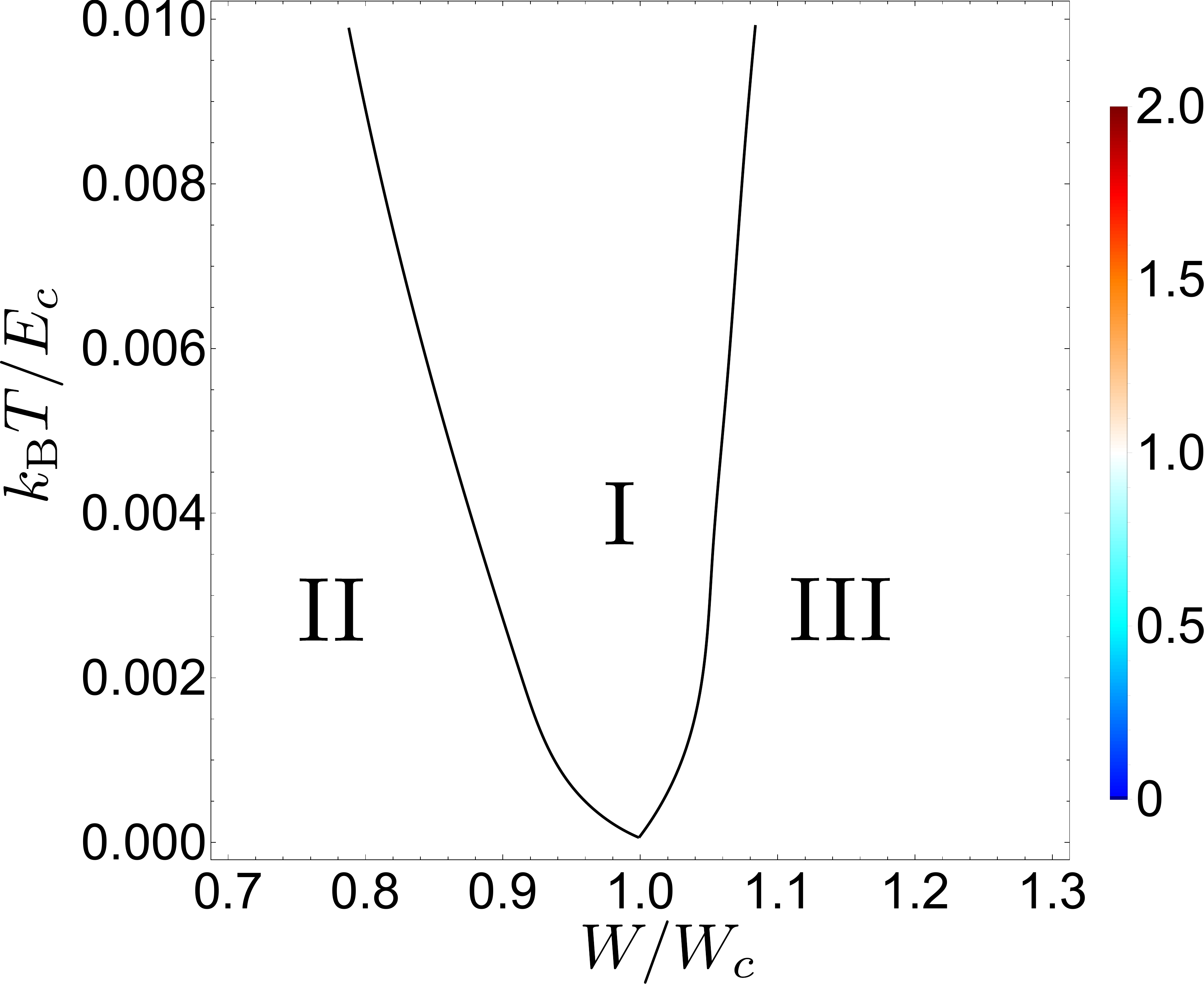}
\centering
\caption{(color online) Temperature-disorder phase diagram obtained by the SCBA. The color code indicates the exponent of $(T_1T)^{-1}$ with respect to the temperature at $E_\textrm{F}=0$, defined as $\kappa(T)$ in the main text. The boundary between three regimes is estimated from $\kappa(T)$ (solid line). }
\label{phase_NMR}
\end{figure}

Our result shows that the temperature dependence of $(T_1T)^{-1}$ reflects the scaling property at the disorder-induced QCP for small $E_\textrm{F}$.
This is clearly illustrated in the temperature-disorder phase diagram (Fig.~\ref{phase_NMR}).
The color code represents the exponent $\kappa(T)$ of the temperature in $(T_1T)^{-1}$ at $E_\textrm{F}=0$, which is estimated as
\begin{align}
\kappa(T)=-\lim_{\Delta T\rightarrow 0}\frac{\log T_1T|_{T+\Delta T}-\log T_1T|_{T}}{\log(T+\Delta T)-\log T}.
\end{align}
Below the QCP, $(T_1T)^{-1}$ is roughly described by the quadratic function with respect to $T$ for a wide range of temperatures (regime II). Only in a narrow region near the QCP, the linear dependence in the temperature is found (regime I). Above the QCP, the finite density of states leads to the constant value of $(T_1T)^{-1}$ (regime III).

\subsection{Asymptotic expressions}
\label{asy_exp}
From the numerical calculation, $(T_1T)^{-1}$ is shown to be linear in the temperature near the QCP.
In order to extract the exact temperature dependence, we derive the expression for the short-ranged scatters without the vertex correction, which is written as
\begin{align}
\frac{\hbar}{T_1k_\textrm{B}T}&=\frac{2\gamma_\textrm{n}^2\mu_0^2E_c}{3 \pi^2c^*}\!\int^\infty_{-\infty}d\omega \! \left(\!\!-\frac{\partial n_{\rm F}(\omega)}{\partial \omega}\!\right)\int^\infty_0\! dQ\, \sigma_\textrm{T}^{(0)}(q),\nonumber
\end{align}
with the expression for $ \sigma_\textrm{T}^{(0)}(q)$ given in Eqn.~(\ref{nov_cond}).
We should note that the scaling of $(T_1T)^{-1}$ at the QCP is not affected by the vertex correction and the additional self-energy correction $Z_{\rm L}^{\rm R}(\omega) $.
This is confirmed by comparing the obtained expression with the numerical results.

For convenience, we shift $Kx$ by $Kx+\frac{Q}{2}$ in $I_n^{\alpha\beta}(K)$ of Eqn.~(\ref{transverse1}) and introduce new integration variables $K'$ and $x'$. 
The transverse component is rewritten as
\begin{align}
\Xi_{\rm T}^{\alpha\beta} (\omega\,; q) =&\int_0^\infty \!\! \frac{K^2dK}{1+K^2} \!\int^\infty_0 \!\! \frac{K'^2dK'}{1+K'^2}\!\left[ {\tilde \Omega}_\alpha {\tilde \Omega}_\beta
I_{0,0}^{\alpha\beta}(K,K')\right. \nonumber\\
&\left. - I_{2,0}^{\alpha\beta}(K,K')-Q I_{1,0}^{\alpha\beta}(K,K') \right]\nonumber\\
&\times\delta(Kx\!-\!K'x'\!+\!Q).
\end{align}
Here, the expressions for the integrals are defined as
\begin{align}
I_{n,m}^{\alpha\beta}(K,K')=&\,K^{n}K^{\prime m} \!\! \int_{-1}^{1} x^n dx  \int_{-1}^{1} x^{\prime m} dx' \nonumber\\
&\times \frac{\delta[K^2(1-x^2)-K^{\prime 2}(1-x'^2)]}{\left( {\tilde \Omega}_\alpha^2 \!-\! K^2 \!\right)\left( {\tilde \Omega}_\beta^2 \!-\! K^{\prime 2 }\!\right)}.
\end{align}
Taking the integration over $Q$, the second term is canceled out by the third term and we obtain
\begin{align}
\int^\infty_0 \!\! dQ\,\Xi_{\rm T}^{\alpha\beta} (\omega\,; q) 
=&\int_0^\infty \!\! \frac{K^2dK}{1+K^2}\int^\infty_K  \!\!\frac{K'^2dK'}{1+K'^2} \nonumber\\
&\times {\tilde \Omega}_\alpha {\tilde \Omega}_\beta I_{0,0}^{\alpha\beta}(K,K'),
\end{align}
where
\begin{align}
I_{0,0}^{\alpha\beta}(K,K')=&\frac{2\textrm{Arcsinh}\left(\frac{K}{\sqrt{K'^2-K^2}}\right)}{K K^{\prime}\left( {\tilde \Omega}_\alpha^2 \!-\! K^2 \!\right)\left( {\tilde \Omega}_\beta^2 \!-\! K^{\prime 2 }\!\right)}.
\end{align}

\subsubsection{Weak disorder }
In the weak disorder regime ($W\ll W_c$), the imaginary part of Green function $G(\boldsymbol{k},\omega)$  is strongly peaked around $\hbar c^*k=\hbar\omega$ with its width proportional to the imaginary part of the self-energy.
Thus, we can simplify the integral by introducing the Dirac delta function $-\pi\,\textrm{Im}[\tilde{\Omega}_\textrm{R}]\delta(K'- \textrm{Re}[\tilde{\Omega}_\textrm{R}])$, where $\tilde{\omega}_\textrm{R}=\omega+\frac{i}{2\tau}$.
Under this approximation, the leading order term is obtained as
\begin{align}
\int^\infty_0 dQ\,\Xi_{\rm T}^{\alpha\beta} (\omega\,; q) 
=&\frac{\pi \textrm{Re}[\tilde{\Omega}_\textrm{R}]\textrm{Im}[\tilde{\Omega}_\textrm{R}]\,{\tilde \Omega}_\alpha {\tilde \Omega}_\beta}{ 16\,({\tilde \Omega}_\beta^2 \!-\! \omega^2\!)} \nonumber\\
&\times\big(\pi+2i \log|\omega|\tau\big)^2\!\!.
\end{align}
We should note that the above integral are convergent in the limit of the infinite momentum cutoff. 
This is expected as the momentum cutoff is not necessary for a clean system. 
Substituting the above expression, the nuclear spin-lattice relaxation rate is derived as
\begin{align}
\frac{\hbar}{T_1k_\textrm{B}T}=\frac{ \gamma_\textrm{n}^2\mu_0^2e^2 }{6\pi^3 c^{*2} }\int^{\infty}_{-\infty} d\omega\left(\!-\frac{\partial n_F(\omega)}{\partial \omega}\right) \omega^2\log |\omega|\tau.
\label{weak_rate2}
\end{align}
For a Weyl electron system without disorder,  $(T_1T)^{-1}\propto T^2\log(2k_\textrm{B}T/\hbar\omega_0)$ with $\omega_0$ denoting the nuclear Larmor frequency~\cite{hirosawa,dora,maebashi1}.
In Eq.~(\ref{weak_rate2}), the nuclear Larmor frequency is replaced with the scattering rate $\frac{1}{2\tau}=\textrm{Im}[\tilde{\omega}_\textrm{R}]=\frac{W}{W_c}\frac{\hbar \omega^2}{E_c}$. 
The equivalent result was obtained in metallic systems~\cite{knigavko}. At $\mu=0$, it is derived as
\begin{eqnarray}
\frac{\hbar}{T_1k_\textrm{B}T}&=&2\pi \left(\frac{ \gamma_\textrm{n}\mu_0e k_\textrm{B}T }{6\pi\hbar c^{*} }\right)^2\Big(\log\frac{2E_c W_c}{k_\textrm{B}T W}-1.05\Big).\quad\quad
\label{weak_rate_mu0}
\end{eqnarray}
Thus, the $T^2$ dependence of $(T_1T)^{-1}$ holds under weak disorder. 
However, the logarithmic term is different from the clean system as the temperature $T$ appears in the denominator.

\subsubsection{QCP }

The SCBA solution at the QCP ($W=W_c$) is given by $\tilde{\omega}_\textrm{R}=\sqrt{\frac{E_c\omega}{2\hbar}}(1+i)$ for small $\omega$. In this case, we cannot simplify the integral by assuming the small imaginary part in the self-energy.
After evaluating the integral over $K$ and $K'$, the leading order term is obtained as
\begin{align}
\int^\infty_0 dQ\,\Xi_{\rm T}^\textrm{RA} (\omega\,; q) 
=&\frac{\pi^2\omega^2\tau^2}{8} \Big[-C_1-C_2 \log\omega \tau \nonumber\\
&+\frac{i\pi}{2}+\log 2\Big]\!,\\
\int^\infty_0 dQ\,\Xi_{\rm T}^\textrm{RR} (\omega\,; q) 
=&\frac{i \pi^2\omega^2\tau^2}{8}  \Big[-C_1-C_2 \log\omega \tau\nonumber\\
&+i\pi-\log 2\Big]\!.
\end{align}
Here, there is a diverging term in the integral so we need a momentum cutoff for convergence.
Since it was not possible to obtain a simple analytical form of the diverging term, the coefficients were estimated as $C_1=1.75$ and $C_2=4.0$ by the numerical fitting. The integral of non-diverging terms is evaluated without the cutoff. 
The nuclear spin-lattice relaxation rate is derived as
\begin{eqnarray}
\frac{\hbar}{T_1k_\textrm{B}T}&=&\frac{e^2\gamma_\textrm{n}^2\mu_0^2E_c}{24\pi^3\hbar c^{*2}}\int^{\infty}_{-\infty} d\omega\left(-\frac{\partial n_\textrm{F}(\omega)}{\partial \omega}\right)\nonumber\\
&\times&|\omega|(\pi+\log2-C_1-C_2\log\omega\tau).
\label{rate_critical2}
\end{eqnarray}
Substituting $\frac{1}{2\tau}=\textrm{Im}[\tilde{\omega}_\textrm{R}]=\sqrt{\frac{E_c \omega}{2\hbar}}$, the expression for $\mu=0$ is obtained as
\begin{align}
\frac{\hbar}{T_1k_\textrm{B}T}=&\frac{\pi  E_ck_\textrm{B}T \log 2}{24}\left(\frac{e\gamma_\textrm{n}\mu_0}{\pi\hbar c^{*} }\right)^2 \\
&\times C_2\left(\log \frac{2E_c}{k_BT}-\frac{2C_1}{C_2}-0.653+\frac{2\pi+2\log 2}{C_2}\right).
\label{critical_rate_mu0}
\end{align}
Therefore, $(T_1T)^{-1}$ is proportional to $T\log \frac{E_c}{k_\textrm{B}T}$ with the constants $C_1$ and $C_2$ dependent on the choice of momentum cutoff. 
In Fig.~\ref{rate2}(a), Eq.~(\ref{critical_rate_mu0}) is plotted with $C_1=4.0$ and $C_2=3.8$ (dashed line), which is in good agreement with the numerical result.
Therefore, the temperature dependence of $(T_1T)^{-1}$ at the QCP is correctly described by Eq.~(\ref{critical_rate_mu0}).

\subsection{Scaling of $(T_1T)^{-1}$ near the QCP }
\label{scaling_section}

The scaling of the conductivity with the system size ($L$) is derived as $\sigma\sim L^{2-d}$ in a $d$-dimensional system~\cite{roy}. This is also obtained from the RG analysis~\cite{sergev_prl}. Since the transverse conductivity is integrated over $q$, we obtain
\begin{eqnarray}
\frac{\hbar}{T_1k_\textrm{B}T}(\delta,\Omega)&=&L^{1-d}G\left(\frac{L}{\delta^{-\nu}},\frac{\Omega}{\delta^{\nu z}}\right)\nonumber\\
&=&\delta^{(d-1)\nu}F\left(\frac{L}{\delta^{-\nu}},\frac{\Omega}{\delta^{\nu z}}\right),
\end{eqnarray}
where $G$ and $F$ are the universal scaling functions, $\delta=W/W_c-1$ and $\Omega=\textrm{max}[\mu(T),k_\textrm{B}T]/E_c$. 

At the QCP ($\delta=0$), the expression for $(T_1T)^{-1}$ should be independent of $\delta$. Thus, it is scaled as
\begin{equation}
\frac{\hbar}{T_1k_\textrm{B}T}(\delta=0,\Omega)\sim\Omega^{\frac{d-1}{z}}.
\label{scale_QCP}
\end{equation}
Above the QCP, the nuclear spin-lattice relaxation rate becomes finite even at a Weyl point. At $\Omega=0$, it is given by
\begin{equation}
\frac{\hbar}{T_1k_\textrm{B}T}(\delta>0,\Omega=0)\sim\delta^{(d-1)\nu}.
\end{equation}

Given $z=2$ and $d=3$ for the SCBA, we obtain $(T_1T)^{-1}\sim T$ at the QCP. Therefore, the scaling analysis is consistent with the present result within the SCBA.
In this special case, the nuclear spin-lattice relaxation rate is proportional to the square of density of states. Generally, there is no simple relationship between the orbital contribution of $(T_1T)^{-1}$ and the density of states at the QCP. 
On the other hand, the critical exponents are obtained as $z=1.5$ and $\nu=1$ in the one-loop RG calculation~\cite{goswami}.  
The numerical calculations of the critical exponents were also performed, yielding $z\approx 1.5$ and $\nu\approx 1$~\cite{kobayashi2,bera}.
Substituting $z=1.5$, we predict $(T_1T)^{-1}\sim T^{\frac{4}{3}}$ at the QCP. Thus, there is no significant difference from the SCBA result except for a slight modification in the exponent. 

\section{Particle-hole asymmetry }
\label{discussion}
While the particle-hole symmetry is conserved in the SCBA, higher order perturbations break this symmetry.
In this section, we discuss the effect of particle-hole asymmetry with the the self-consistent $T$-matrix approximation (SCTA), which takes account of characteristic higher order corrections with respect to impurity potential $u_{ab}$.

\begin{figure}
\includegraphics[width=70mm]{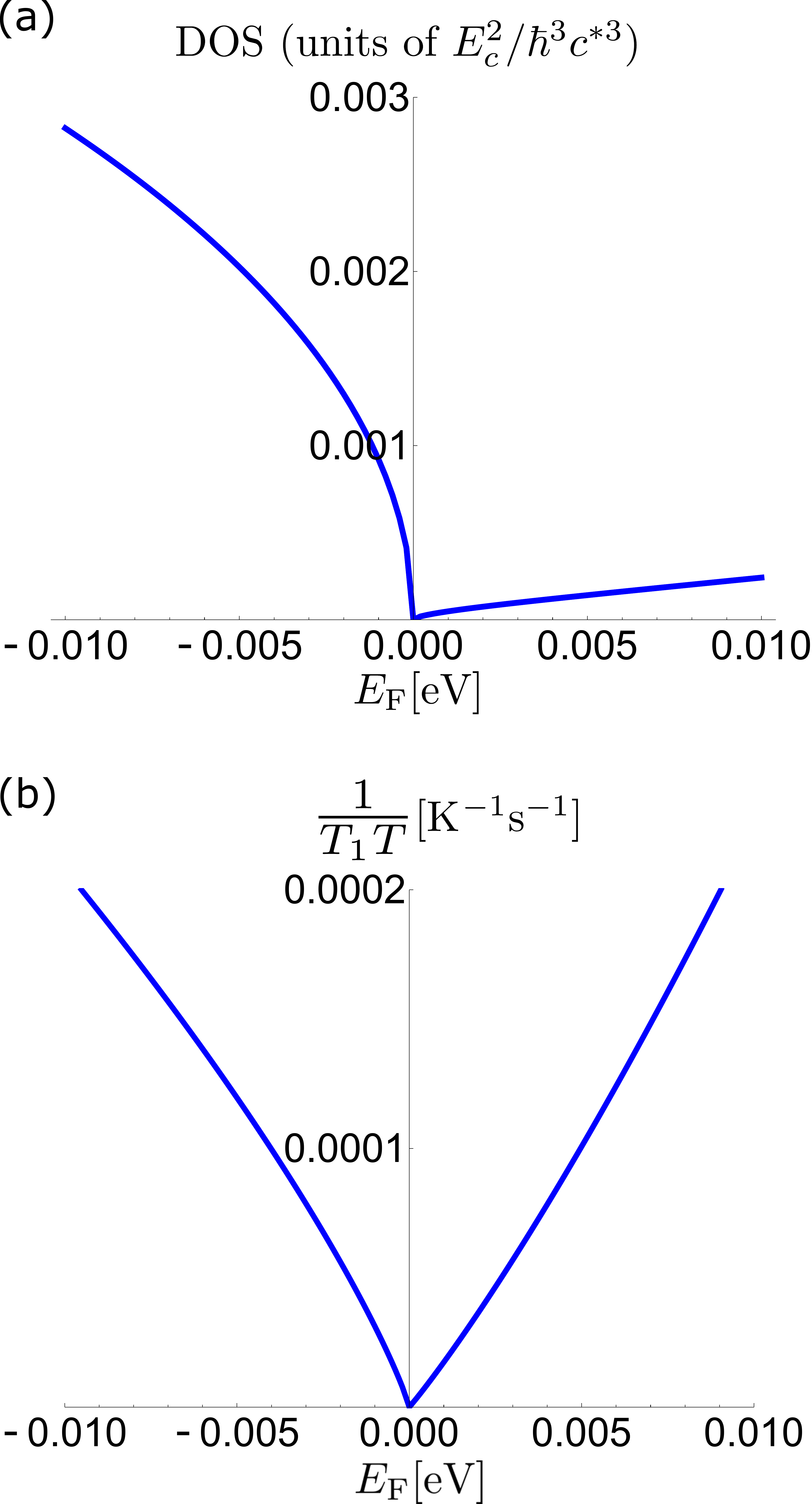}
\centering
\caption{ (a) The density of states and (b) the nuclear spin-lattice relaxation rate $(T_1T)^{-1}$ at $T=0$ are plotted against the Fermi energy $E_\textrm{F}$ at the QCP. 
For both plots, the impurity potential is fixed at $\bar{u}\approx 0.6$. The other parameters are the same as in Fig.~\ref{rate1}. 
}
\label{fig:tmatdos}
\end{figure}

\subsection{Critical exponents in the SCTA}

The self-consistent equation for the SCTA is given by
\begin{equation}
\Sigma^\textrm{R}_a(\omega)=\frac{n_i|u_{aa}|}{\hbar}\left(1-\sum_{b=\mathcal{L},\mathcal{R}}\frac{|u_{ab}|}{\hbar}\sum_\textbf{k}G_b^\textrm{R}(\textbf{k},\omega)\right)^{-1}\!\!\!\!,
\end{equation}
where the subscript $a$ denotes the chirality. 
We define the impurity concentration and impurity potential as $\bar{n}_i=a_c^3n_i$ and $\bar{u}=(|u_{\mathcal{LL}}|+|u_{\mathcal{RL}}|)/a_c^3E_c$, where the effective lattice constant $a_c=2\pi/k_c$ is introduced.
Similarly to the SCBA, the self-energy has an identity matrix element satisfying 
\begin{equation}
\Sigma_{\text{SCTA}}(\omega)=\frac{\bar{n}_i\bar{u}}{\hbar(1-4\pi\bar{u}\tilde{\Omega}^\textrm{R}f(\omega))},
\label{scta1}
\end{equation}
with the expression of $f(\omega)$ given in Eq.~(\ref{func_F}) and $\tilde{\Omega}_\textrm{R}=\hbar\tilde{\omega}_\textrm{R}/E_c$. 
Here, the chemical potential is shifted so that $\tilde{\omega}_\textrm{R}=\omega+(\mu+\bar{n}_i\bar{u}E_c)/\hbar-\Sigma_\textrm{I}^\textrm{R}(\omega)$.
We should note that the impurity concentration $\bar{n}_i$ needs to be small for justifying the SCTA.
For long-ranged disorder, the impurity concentration is multiplied by $\frac{1}{2}$.
The self-consistent solution is derived as
\begin{align}
\tilde{\Omega}^\textrm{R}=&\frac{1}{2}\Big(\Omega+\frac{\delta}{(2\pi\bar{u}^2-i)} \nonumber\\
&\pm\frac{\sqrt{4\Omega(2\pi\bar{u}^2-i)+\big[\Omega(2\pi\bar{u}^2-i)+\delta\big]^2}}{(2\pi\bar{u}^2-i)}\Big),
\label{scta}
\end{align}
where $\delta=\bar{n}_i/\bar{n}_c-1$.
Similarly to the SCBA, the critical impurity concentration is defined at which the imaginary part of the self-energy becomes finite. At the QCP ($\bar{n}_c=W_c/4\pi\bar{u}^2$) , the solution for $\Omega\ll 1$ is given by
\begin{align}
\tilde{\Omega}^\textrm{R}= \sqrt{\frac{2\Omega}{-2i+\pi \bar{u}}}.
\end{align}
Above the QCP at $\Omega=0$, it is given by
\begin{align}
\tilde{\Omega}^\textrm{R}=\frac{2\delta(\pi\bar{u}+2i)}{4+\pi \bar{u}^2}.
\label{scta_QCP}
\end{align}
Thus, the critical exponents are identical with the SCBA ($z=2,\,\nu=1$). In the limit of $\bar{u}\rightarrow 0$, Eq.~(\ref{scta}) turns into the solution for the SCBA.

Using the solution for the SCTA, we study the critical behavior under the strong impurity potential. 
The impurity potential is fixed at $\bar{u}\approx0.6$, giving the critical impurity concentration $\bar{n}_c\approx0.14$. 
In Fig.~\ref{fig:tmatdos}(a), the density of states against the Fermi energy is calculated at the QCP.
The difference from the SCBA is that the density of states is not symmetric about the Weyl point.
Thus, the particle-hole symmetry is broken under the strong impurity potential, although the square root singularity is still obtained at the QCP.
Above the QCP, the density of states at the Weyl point is suppressed by $\bar{u}^{2}$ in the denominator of Eq.~(\ref{scta_QCP}).
We also confirm that the the nuclear spin-lattice relaxation rate is linearly proportional to the temperature at the QCP as in the SCBA.
This is illustrated in Fig.~\ref{fig:tmatdos}(b), showing the linear dependence of $(T_1T)^{-1}$ with respect to the Fermi energy at $T=0$ for small $E_\textrm{F}$. 
Here, we ignore the vertex correction and the $q$ dependent self-energy for simplicity.

\subsection{Enhanced upturn in ($T_1T$)$^{-1}$}

In the previous section, we show that the higher order contributions in impurity scattering result in the particle-hole symmetry breaking.
This is particularly important at finite temperatures, as it modifies the temperature dependence of the chemical potential.
We should note that the particle-hole symmetry is recovered in the unitary limit as we can ignore unity in the denominator of Eq.~(\ref{scta1}) in the limit of $\bar{u}\rightarrow \infty$ \cite{ostrovsky}.

\begin{figure}
\includegraphics[width=75mm]{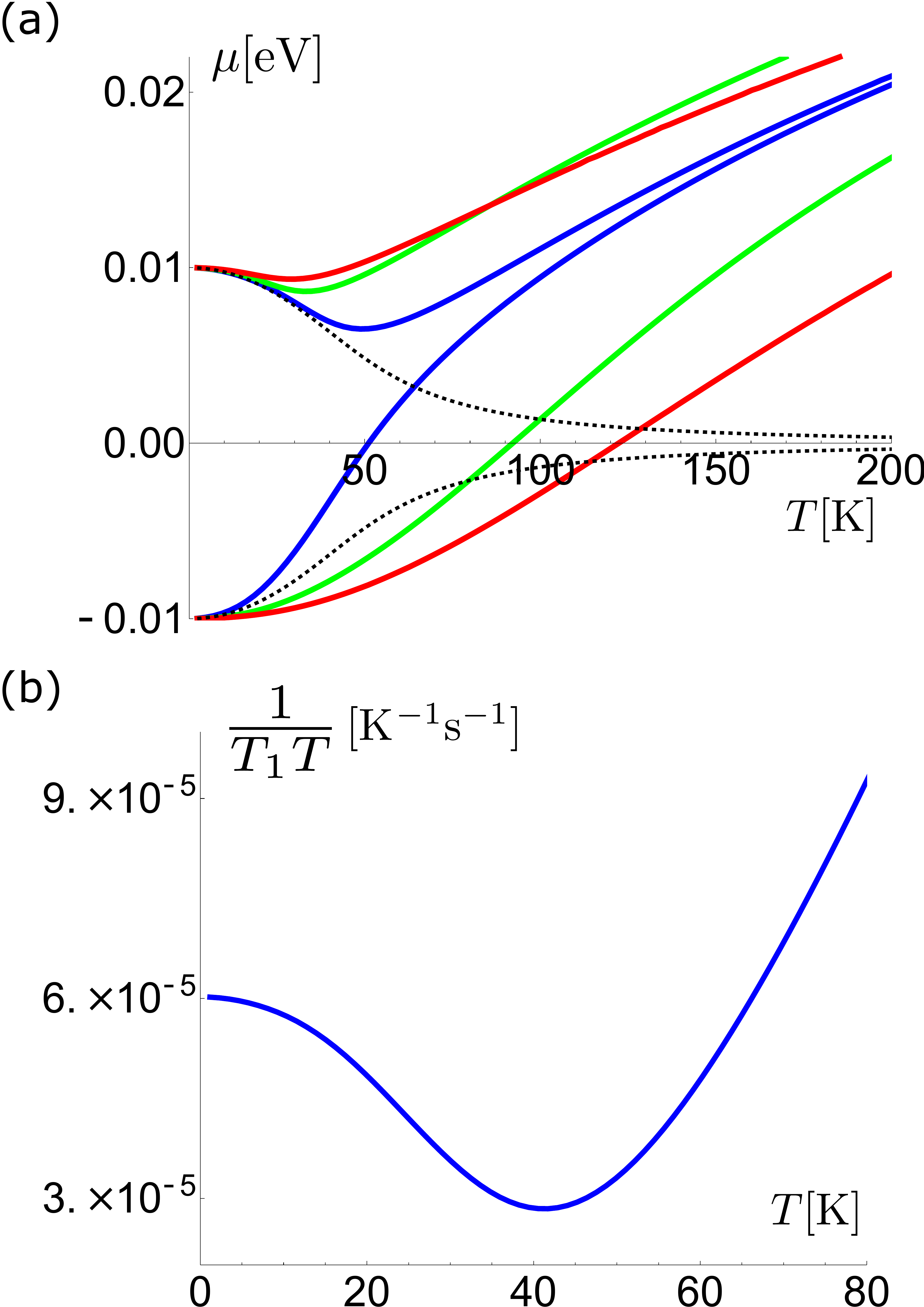}
\centering
\caption{(color online) 
(a) Chemical potential against temperatures within the SCTA for $E_\textrm{F}=\pm 10$ meV.
The impurity concentration is set at $\bar{n}_i/\bar{n}_c= 0.3$ (blue), $ 0.6$ (green), $ 0.9$ (red) from the Weyl point.  The dashed lines show the result of clean systems. 
(b) Plot of $(T_1T)^{-1}$ at low temperatures for $\bar{n}_i/\bar{n}_c=0.3$ and $E_\textrm{F}=-10$ meV. 
}
\label{fig:tmatrate}
\end{figure}

Within the SCBA that preserves the particle-hole symmetry, the chemical potential moves towards Weyl points.
Under strong impurity potential, this behavior is significantly modified in order to compensate for the imbalance of the density of states. 
In Fig.~\ref{fig:tmatrate}(a), the chemical potential is plotted against temperatures at $E_\textrm{F}=\pm 10$ meV.
As the density of states is larger for $E_\textrm{F}<0$ (Fig.~\ref{fig:tmatdos}(a)), the chemical potential tends to increase with the temperature.
This results in an upturn of $\mu(T)$ for $E_\textrm{F}>0$ and a reverse of the sign for $E_\textrm{F}<0$.
The important point is that the shift of $\mu(T)$ at low temperatures is strongly enhanced for $E_\textrm{F}<0$ compared to the clean limit. 

As discussed in Section \ref{sec: numerical_rate}, the nuclear spin-lattice relaxation rate $(T_1T)^{-1}$ shows the upturn at $k_\textrm{B}T\sim E_\textrm{F}/2$ due to the shift of chemical potential towards the Weyl point.
Under the strong impurity potential, the overshooting of $\mu(T)$ above the Weyl point may happen as a result of the particle-hole asymmetry.
This leads to the enhancement in the upturn of  $(T_1T)^{-1}$.
In Fig.~\ref{fig:tmatrate}(b), the low temperature behavior of $(T_1T)^{-1}$ is shown at $E_\textrm{F}=-10$ meV and  $\bar{n}_i/\bar{n}_c= 0.3$.
We find that it drops by half from $T=0$ to $T\sim 40$ K before the uprising of the $T^2$ term, which is much larger than the upturn of  $(T_1T)^{-1}$ in a clean system.

In the recent experiment, the upturn in $(T_1T)^{-1}$ was observed in TaP~\cite{yasuoka}.
One discrepancy from the theory is that the observed upturn was much larger than the theoretical result~\cite{okvatovity}.
As discussed in Ref.~\cite{okvatovity}, it is possible to explain this behavior by introducing the phenomenological form of chemical potential $\mu(T)$.
In addition, we point out that the upturn is amplified by the asymmetry in the density of states about the Weyl node.
Under the strong impurity potential, even a small amount of impurities results in the particle-hole asymmetry and enhancement of the upturn in $(T_1T)^{-1}$, which could explain the experimental result.

\section{Conclusion}

We have studied the nuclear spin-lattice relaxation rate due to orbital currents in disordered Weyl fermion systems, employing the self-consistent Born approximation (SCBA).
In this work, two types of the disorder potential was considered, namely the short-ranged (intervalley and intravalley scattering) and long ranged scatters (only intravalley scattering).
For both cases, it shows the critical behavior with the critical exponents $z=2$ and $\nu=1$ in the SCBA.

The orbital contribution of the nuclear spin-lattice relaxation rate is determined by the transverse conductivity, whose wavevector dependence was investigated under disorder. 
The vertex correction was obtained in a gauge invariant manner for general wavevector $q$, using  the conserving approximation for the SCBA. 
As shown in Fig.~\ref{sigmaq},  the vertex correction has a negative contribution for the long-ranged scatters, while it vanishes at $q= 0$ for the short-ranged scatters.

Our main result is the scaling relation of the nuclear spin-lattice relaxation near the disorder-induced QCP.
As shown in Fig.~\ref{phase_NMR}, we classified three different regimes from the temperature dependence of $(T_1T)^{-1}$.
For each regime, the scaling of the nuclear spin-lattice relaxation rate is summarized in Table~\ref{table1}.
Within the SCBA, we obtained the asymptotic expression at the QCP as $(T_1T)^{-1}\sim T\log(1/T)$.
Although the critical exponents from the one-loop renormalization group analysis predicts the exponent in $(T_1T)^{-1}$ slightly greater than the SCBA result~\cite{goswami}, our result provides a good physical picture near the disorder-induced QCP in Weyl fermion systems.

In addition, we discussed the effect of the particle-hole asymmetry, employing the self-consistent T-matrix approximation (SCTA). In the regime II under the strong impurity potential, the temperature dependence of chemical potential is significantly modified from the particle-hole symmetric systems. As a result, the low temperature upturn in $(T_1T)^{-1}$ becomes enhanced, which could explain the large upturn observed in the experiment~\cite{yasuoka}. 

\begin{table}[h]
\begin{center}
\def\arraystretch{1.2}
\begin{tabular}{l|cc}
 &  ~SCBA \& SCTA~ &~Scaling ansatz~\\ \hline
 Regime I& $E\log 1/E$& $E^{\frac{2}{z}}$\\ 
 Regime II& $E^2\log 1/E$ & $E^2$ \\ 
Regime III&$\delta^{2}$ & $\delta^{2\nu}$
\end{tabular}
\caption{The scaling relation of the nuclear spin-lattice relaxation rate $(T_1T)^{-1}$. We denote the characteristic energy as $E=\max[k_\textrm{B}T,\mu(T)]$ and the impurity strength $\delta=W/W_c-1$. }
\label{table1}
\end{center}
\end{table}

\appendix

\section{Approximate solution for the self-energy correction}
\label{approximate solution}
Here, we derive the $q$ dependent self-energy to conserve the gauge invariance. 
From Eqs.~(\ref{self-energy correction}) and (\ref{SolSCBA'}), 
\begin{align}
\Sigma_\textrm{I}^\textrm{R}(\omega;q/2)&=W\tilde{\omega}_\textrm{R}f(\omega;q/2),
\label{A1:scba1}
\\
Z_{\rm L}^{\rm R}(\omega;q/2)-1&=W\frac{2g(\omega;q/2)+\tilde{Q}_\textrm{R}f(\omega;q/2)}{2Q},
\label{A1:scba2}
\end{align}
where $\tilde{\omega}_\textrm{R}=\omega+\mu/\hbar-\Sigma_\textrm{I}^\textrm{R}(\omega;q/2)$, $Q=q/k_c$, and $\tilde{Q}_\textrm{R}=Z_\textrm{L}^{\rm R}(\omega;q/2)\frac{q}{k_c}$. In order to solve the above self-consistent equations, we expand $f(\omega;q/2)$ and $g(\omega;q/2)$ for small $q$. 
This approximation is justified as we are interested in the integration of the transverse conductivity over $q$, which converges for $q\ll k_c$ under weak disorder.
As the disorder strength is increased, the transverse conductivity becomes constant with respect to $q$. Thus, the $q$ dependence in the self-energy does not affect the integral for $q \sim k_c$.
The dimensionless functions $f(\omega;q/2)$ and $g(\omega;q/2)$ for $ q\ll k_c$ are given as
\begin{align}
f(\omega;q/2)=&\int^{\infty}_0\frac{ K^2dK}{K^2+1}\int^{1}_{-1}\frac{dx}{2}\frac{1}{\tilde{\Omega}_\textrm{R}^2-K^2-K\tilde{Q}_\textrm{R}x-\frac{\tilde{Q}_\textrm{R}^2}{4}} \nonumber\\
=& -\frac{\pi}{2}\frac{1+i\tilde{\Omega}_\textrm{R}}{1+\tilde{\Omega}_\textrm{R}^2}+\textrm{O}(\tilde{Q}_\textrm{R}^2),
\end{align}
\begin{align}
g(\omega;q/2)=&\int^{\infty}_0\frac{ K^2dK}{K^2+1}\int^{1}_{-1}\frac{dx}{2}\frac{Kx}{\tilde{\Omega}_\textrm{R}^2-K^2-K\tilde{Q}_\textrm{R}x-\frac{\tilde{Q}_\textrm{R}^2}{4}} \nonumber\\
=&\frac{\pi\tilde{Q}_\textrm{R}}{12}\frac{2+i\tilde{\Omega}_\textrm{R}(3+\tilde{\Omega}_\textrm{R}^2)}{(1+\tilde{\Omega}_\textrm{R}^2)^2}+\textrm{O}(\tilde{Q}_\textrm{R}^3),
\end{align}
where $\tilde{\Omega}_\textrm{R}=\hbar\tilde{\omega}_\textrm{R}/E_c$.
The solution of Eqs.~(\ref{A1:scba1}) and (\ref{A1:scba2}) are given in Eqs.~(\ref{sol_1}) and (\ref{sol_2}).
In this approximation, the identity matrix element $\Sigma_\textrm{I}^\textrm{R}$ is unchanged.

\begin{acknowledgments}
We would like to thank I. Tateishi, V. K$\ddot{\textrm{o}}$nye, H. Matsuura, and H. Yasuoka for helpful comments and discussions. 
This work was supported by Grants-in-Aid for Scientific Research from the Japan Society for the Promotion of Science (Grant Nos. JP18H01162 and JP18K03482).
T. H. is supported by Japan Society for the Promotion of Science through Program for Leading Graduate Schools (MERIT) and JSPS KAKENHI (Grant No. 16J07110).   
\end{acknowledgments}

\end{document}